\documentclass[printer]{aa}

\bibpunct{(}{)}{;}{a}{}{,}

\usepackage{graphicx}
\usepackage{txfonts}
\usepackage{soul}
\usepackage{txfonts}
\usepackage{color}
\usepackage{ulem}

\newcommand{\hii}{H\textsc{ii}}
\def\ks{km s$^{-1}$}

\def\s{$^{\prime\prime}$}

\def\cm3{cm$^{-3}$}

\def\2{$^{12}$CO}
\def\3{$^{13}$CO}
\def\8{C$^{18}$O}
\def\msol{M$_\odot$}

\def\cm2{cm$^{-2}$}

\begin{document}

\title{Revisiting G29.862$-$0.0044: a jet cavity disrupted by an outflow in a likely young stellar object wide binary system}

\author{S. Paron \inst{1}
\and N. C. Martinez \inst{1,2}
\and M. E. Ortega\inst{1}
\and D. Mast\inst{3}
\and A. Petriella\inst{1}
\and L. Supán\inst{1}
\and C. Fariña\inst{4,5}
}
\institute{CONICET - Universidad de Buenos Aires. Instituto de Astronom\'{\i}a y F\'{\i}sica del Espacio
             CC 67, Suc. 28, 1428 Buenos Aires, Argentina\\
             \email{sparon@iafe.uba.ar}
\and Universidad de Buenos Aires, Facultad de Ciencias Exactas y Naturales, Departamento de Física. Buenos Aires, Argentina
\and Observatorio Astronómico, Universidad Nacional de Córdoba, Laprida 854, X5000BGR Córdoba, Argentina
\and Isaac Newton Group of Telescopes, E38700, La Palma, Spain
\and Instituto de Astrof\'{\i}sica de Canarias (IAC) and Universidad de La Laguna, Dpto. Astrof\'{\i}sica, Spain
}

\offprints{N. C. Martinez}

   \date{Received <date>; Accepted <date>}

\abstract{}{A few years ago, we investigated MYSO G29.862$-$0.0044 (YSO-G29), an intriguing star-forming region at a distance of 6.2 kpc. Although the typical disc-jet scenario was proposed to explain the observations, it remained far from conclusive. We wonder if the puzzling observed near-IR features are produced by only one source, or it is due to confusion generated by an unresolved system of YSOs. Unveiling this issue is important for a better understanding of the star-forming processes.}
{YSO-G29 was analysed using new observations at near-IR from Gemini-NIFS, at radio continuum (10 GHz) from Jansky Very Large Array (JVLA), and new continuum (1.3 mm) and molecular line data from the Atacama Large Millimeter Array (ALMA).}
{
The near-IR observations allowed us to detect emission of H$_{2}$ 1–0 S(1) and Br$\gamma$ lines in YSO-G29, which are compatible with excitation and ionization from UV radiation propagating in a highly perturbed ambient. In addition, some evidence of H$_{2}$ excitation by collisions were found. The ALMA data show the presence of a conspicuous and collimated molecular outflow propagating southwards, while to the north, an extended molecular feature perfectly surrounded by the Ks near-IR emission appears. The continuum emission at 1.3 mm allowed us to better resolve the molecular cores, one of which stands out due to its high temperatures and rich chemical composition. From the JVLA observations, we discovered a compact radio continuum source, a likely compact \hii~region or an ionised jet of a massive protostar, located at $\sim$0\farcs7 ($\sim$0.02 pc) from the main millimetre core. In this way, we propose a YSO wide binary system.}
{We can explain the nature of the intriguing near-IR features previously observed: cone-like structures produced by jets/winds of one of the components of the binary system that cleared out the surroundings were disrupted by a molecular outflow probably from the other component. These results complete the picture of what is happening in YSO-G29, and reveal a phenomenon that should be considered when investigating
massive star-forming regions.  }

\titlerunning{G29.862$-$0.0044 with Gemini-NIFS, ALMA, and JVLA observations}
\authorrunning{S. Paron et al.}

\keywords{Stars: formation --- Stars: protostars --- ISM: jets and outflows --- ISM: molecules}

\maketitle

\section{Introduction}

Massive stars (canonically with masses and luminosities above 8 \msol~and 10$^{4}$ L$_{\odot}$) form deeply embedded within molecular cores in molecular clumps (see e.g. \citealt{kumar20,moscadelli21,beuth25}). These are
places with very high visual absorption due to the presence of abundant interstellar dust. 
Moreover, since that massive stars tend to form in clusters, the regions in which they form are very confused.
These issues make it difficult to obtain useful observational data on individual massive young
stellar objects (MYSOs), and hence, our understanding of the physics and chemistry of massive star formation 
remains incomplete. Therefore, efforts to observe such regions in the multiwavelength regime are fundamental. 

The molecular cores in which the stars form, called hot molecular cores (HMCs), are among the chemically richest regions in the interstellar medium (ISM) 
(e.g. \citealt{herbst09,bonfand19}), and the star-forming processes strongly influence the chemistry of these environments \citep{jorgen20}. 
The observation of molecular lines and the study of their emission and chemistry are essential for characterizing the physical and chemical conditions of the gas and for constraining the evolutionary stage of a star-forming region. 

To investigate high-mass star formation is appropriate to carry out large surveys of MYSOs, molecular outflows, and extended near-IR H$_{2}$ 
emission associated with high-mass young stars (\citealt{lumsden13,navarete15,caratti15,maud15a,maud15b,yang18}) that provide important information from a statistical point of view.
Additionally, we point out that selecting particular objects to perform dedicated observations for deeper and more detailed studies is also essential.
Studies of particular sources (see, e.g.,  \citealt{gredel06,fedriani18,fedriani19,ferrero22,paron22}) in which the observations, mainly at near-IR and (sub)millimeter wavelengths, are analysed in depth, 
are extremely valuable for shedding light on the formation processes of MYSOs and the physical and chemical characterization of these interstellar environments.

A few years ago we investigated MYSO G29.862$-$0.0044 (hereafter YSO-G29) \citep{areal20,areal21}. Using near-IR data from Gemini-NIRI, millimetre data from the James Clerk Maxwell Telescope, observations from the Atacama Submillimeter Telescope Experiment, and some Atacama Large Millimeter Array (ALMA) data, we performed a multi-spatial scale analysis of this a source, located at the kinematic near distance of about 6.2 kpc and associated with the star-forming region G29.96-0.02 (W43-South, \citealt{carl13}). More recently, in \citet{martinez24}, also using ALMA data we briefly reported the chemical complexity of the region, confirming the hot core nature of the molecular structure related to YSO-G29.

\begin{figure}
    \centering
	\includegraphics[width=8.8cm]{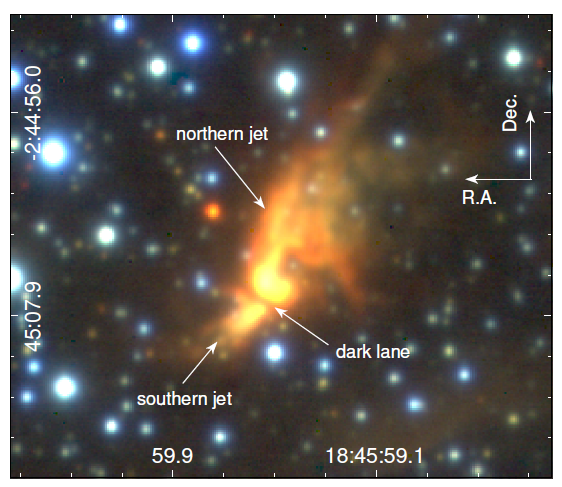}
    \caption{Three-colour image of a 55\s$\times$45\s~region towards YSO-G29 obtained with Gemini-NIRI in our previous work, showing the J, H, and Ks broad-bands emission in blue, green, and red, respectively. The indicated features correspond to the disc-jet system scenario discussed in \citet{areal20}.}
    \label{present}
\end{figure}

Particularly, in \citet{areal20}, we found that YSO-G29 exhibits a conspicuous asymmetric morphology
at both clump and core spatial scales. We proposed a scenario in which the YSO jet has flowed more freely
towards the north consistent with the direction of a redshifted molecular outflow observed at low angular resolution, generating striking and extended features in the near-IR (see Fig.\,\ref{present}). Such features, commonly observed at these wavelengths, are due to cavities cleared out in the circumstellar material by the action of jets and winds \citep{sanna19,paron16,paron13,bik05,bik06,reip00}.

Although a typical disc-jet scenario was proposed for YSO-G29 in \citet{areal20}—with the disc represented by a dark lane (Fig.\,\ref{present})—we now question whether the observed near-IR features arise from a single source or from confusion caused by an unresolved system of YSOs and/or it is due to a complex protostar(s) dynamics. Unveiling this issue is important not only for interpreting this intriguing source but also for improving our understanding of the underlying star-forming processes. 

Hence, we decided to carry out new observations to investigate the nature of this region and its features in greater detail. We used the Jansky Very Large Array (JVLA) to map the radio continuum emission and the NIFS instrument at Gemini-North to perform near-IR spectroscopy. In addition, we analysed more recent ALMA data with higher angular resolution than in our previous studies to examine the molecular gas distribution associated with the striking observed near-IR features. Notably, one spectral window of these new ALMA data includes the $^{12}$CO J=2--1 line and some of its isotopes, which are well suited for investigating potential molecular outflows extending on small spatial scales.

\section{Observations and data}
\label{obs}

This section outlines the datasets used. We begin by describing the new observations obtained with Gemini-NIFS and the JVLA, and subsequently present the archival ALMA data. 

\subsection{Near-IR observations using Gemini-NIFS}

\begin{figure}[h]
    \centering
    \includegraphics[width=8cm]{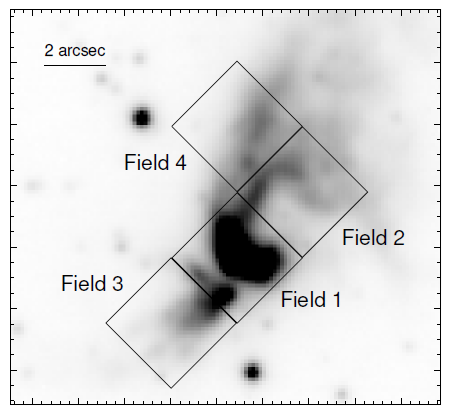}
    \caption{Fields observed using Gemini-NIFS superimposed over the Ks image of YSO-G29 obtained with Gemini-NIRI.}
    \label{nifsObs}
\end{figure}

Near-IR integral field spectroscopic observations were carried out using the Near-infrared Integral Field Spectrograph \citep[NIFS,][]{McGregor2003} mounted on Gemini North, during the first semester of 2022 (project: GN-2022A-Q-125, PI: S. Paron). Four fields towards YSO-G29 were observed (see Fig.\,\ref{nifsObs} and Table\,\ref{nifsTab}) covering almost the entire near-IR structure analysed in \citet{areal20}. The observations were performed in seeing-limited mode using Peripheral WaveFront Sensor 2 (PWFS2) for guiding, rather than with adaptive optics correction via the ALTAIR system, due to the difficulty in finding suitable guide stars for this object. The K\_G5605 grating (central wavelength: $2.20\,\mu$m, 
spectral resolution 5290) together with the HK\_G0603 filter (central wavelength: $2.16\,\mu$m) 
were used to observe the four fields in the K-band only. The observations followed the Object-Sky-Object dithering sequence, with off-source sky positions at 90 arcsec. The spectra are centred at 2.2 $\mu$m covering a spectral range of 2.009--2.435 $\mu$m. The spectral resolution ranges from 2.4 to 4.0 \AA, as determined from the full width at half maximum (FWHM) measured in the ArXe lamp lines used for wavelength calibration. 

In seeing-limited mode the angular resolution is in the range of $0\farcs22-0\farcs26$, derived from the FWHM of the flux distribution of telluric standard stars, corresponding to $\approx0.007$ pc (1444 au) at the distance of YSO-G29. 
The standard NIFS tasks included in the Gemini IRAF package v1.16\footnote{https://www.gemini.edu/sciops/data/software/gemini\_v1161\_rev.txt} were used for data reduction. The procedure included shifting to the MDF file, flat-fielding, sky subtraction, wavelength calibration, and correction for spatial curvature and spectral distortion. Telluric correction of the fields was performed when appropriate by observing two telluric standards: HD 164222 and HD 183596. The telluric correction process included fitting the hydrogen absorption lines of the stellar spectrum, and fitting a synthetic blackbody spectrum to recover the correct shape of the spectral distribution. These telluric stars were also used to flux calibrate the datacubes. The main observational parameters are summarized in Table\,\ref{datanifs}. 

Finally, data cubes were created for each field with a Field-of-View (FoV) of $3\arcsec\times3\arcsec$ and an angular sampling of $0\farcs05\times0\farcs05$. For the analysis of the datacubes DS9\footnote{https://sites.google.com/cfa.harvard.edu/saoimageds9} and QFITSview\footnote{https://www.mpe.mpg.de/$\sim$ott/QFitsView/} software were used. 
For each emission line detected in the datacube, this analysis consists of fitting a continuum to each spaxel and then integrating over the full line profile. With this information, we were able to construct the line emission maps.

\begin{table}[h]
\centering
\caption{Observed fields using NIFS at Gemini North.}
 \begin{tabular}{ccccc}
\hline
\hline
Field & RA          & Dec          & Exp. Time  & PA \\
      &             &              &  (sec)    &  (deg) \\
\hline
1     & 18:45:59.55 & -02:45:06.35 &  180       & 45  \\
2     & 18:45:59.41 & -02:45:04.20 &  360       & 45   \\
3     & 18:45:59.69 & -02:45:08.46 &  360       & 45    \\
4     & 18:45:59.55 & -02:45:02.06 &  360       & 45     \\
\hline
\label{nifsTab}
\end{tabular}
\end{table}

\begin{table}[h]
\centering
\caption{NIFS-Gemini observational parameters.}
\label{datanifs}
\begin{tabular}{ll}
\hline
\hline
  Project &  2022A-Q-125  \\   
  Obs. dates & 2022-05-15/17 \\
  Observing Mode & Seeing-limited \\
  Central wavelength & $2.2\,\mu$m \\  
  Spectral range &  2.009 -- 2.435 $\mu$m    \\ 
  Spectral resolution & 2.4 to 4.0 \AA \\
  Angular resolution &   0\farcs22 -- 0\farcs26   \\
  FOV of each region  &  $3\arcsec\times3\arcsec$          \\
                       
\hline
\end{tabular}
\end{table}

\subsection{Radio observations using JVLA}

We observed YSO-G29 with the JVLA at X frequency band ($10~\mathrm{GHz}$; $\lambda \sim$ 3 cm) in the A-array configuration (project ID: 22A$-$063, PI: M. Ortega), which provides the best angular resolution of the interferometer (sub arcsecond resolutions) in a field of view (FOV) of about 4 arcmin. We used the full $8-12~\mathrm{GHz}$ bandwidth with 32 spectral windows and full polarization. 3C 286 and J1832$-$105 were used as primary and secondary calibrators, respectively. 
The secondary calibrator is catalogued as C in the VLA calibrator source list\footnote{https://science.nrao.edu/facilities/vla/observing/callist}, which correspond to a positional accuracy between 0.01 and 0.15 arcsecs. The total on-source integration time was 30 min. We obtained calibrated visibilities from the CASA pipeline version 5.6.2-2.el7. Additional flagging of RFI was manually performed before imaging. The image was obtained using the tclean task and Briggs weighting. We set ROBUST = 0.25 to reduce sidelobe effects from the nearby complex of bright \hii~regions located at $\sim4^{\prime}$ to the northwest of YSO-G29. The final image has a synthesized beam of $0\farcs42 \times 0\farcs17$ (FWHM) and positional angle (P.A.) of $-51.6^{\circ}$, and a mean noise (rms) $\sigma_{rms}=6~\mathrm{\mu Jy\,beam^{-1}}$ (SNR $\geq$ 30). The main observational parameters are summarized in Table\,\ref{datavla}. 

Additionally, we generated an image of the secondary calibrator J1832$-$105 (the phase calibrator) to assess the astrometric error in the final image. This source exhibits a clear point-like morphology, and its measured position was consistent with the coordinates given by the VLA calibrator source list, showing an offset of less than 0$\farcs$1.

\begin{table}[h]
\centering
\caption{ JVLA observational parameters.}
\label{datavla}
\begin{tabular}{ll}
\hline
\hline
  Project &  22A$-$063  \\               
  Obs. date & 2022-04-15 \\
  Band    &  X   \\
  Freq. range &   7.97 -- 12.02 GHz    \\ 
  Beam size &   0\farcs42 $\times$ 0\farcs17    \\
  rms level & 6.0 $\mu$Jy beam$^{-1}$    \\ 
  Max.reco.scale &  5\farcs3\   \\ 
  FOV   &  4\farcm2              \\
                       
\hline
\end{tabular}
\end{table}

\subsection{ALMA data}
\label{alma data}

The continuum emission image and data cubes from Project 2021.1.00311.S (PI: Liu, Sheng-Yuan) at Band 6 were retrieved from the ALMA Science Archive\footnote{http://almascience.eso.org/aq/}. The single pointing observations for the target were carried out using the 34.5/226.8 telescope configuration in the 12~m array. Table\,\ref{data_info} shows the main ALMA data parameters. The data is QA2 quality level, which assures a reliable calibration for a ``science ready'' data. The astrometric accuracy of the data is about 3 milliarcsec (mas). 

\begin{table}[h]
\centering
\caption{Used ALMA data.}
\label{data_info}
\begin{tabular}{ll}
\hline
\hline
  Project &  2021.1.00311.S   \\               
  Obs. date & 2022-08-13 \\
  Band    &  6   \\
  Freq. range &   216.61 -- 233.72 GHz    \\ 
  Beam size &   0\farcs3     \\
  Line sens.(10~\ks) & 2.8 mJy beam$^{-1}$    \\ 
  $\Delta\nu$ & 1.1 MHz          \\ 
  $\Delta$v   &  1.2 \ks        \\
  Max.reco.scale &  4\farcs1\   \\ 
  FOV   &  25\farcs8              \\
                       
\hline
\end{tabular}
\end{table}

The continuum map was obtained averaging the continuum emission of each of the four spectral windows and was corrected for primary beam. 
The continuum map has an rms noise level of about 0.2 mJy beam$^{-1}$.
The beam size of the observations (0.3 arcsec) provides
a spatial resolution of about 0.009 pc ($\sim$1800 au) at the distance of 6.2 kpc.

\section{Results}

We present here the results from our analysis of YSO-G29 across different wavelengths.

\subsection{The near-IR IFU spectroscopy}

\begin{figure}[h]
   \centering
       \includegraphics[width=8.5cm]{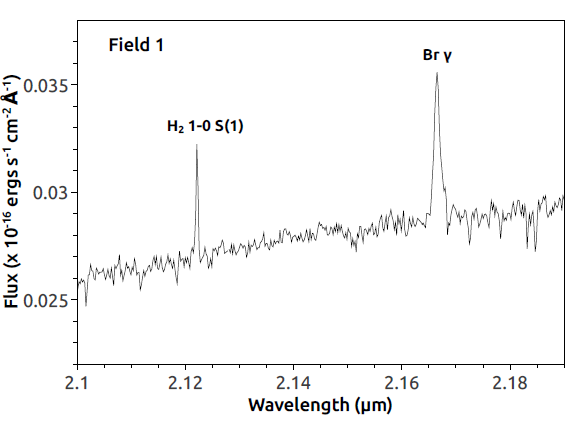}
   \caption{Average spectrum obtained towards Field\,1 presented as an example to show the detected near-IR lines.}
   \label{f1spect}
\end{figure}

\begin{figure}[h]
    \centering
       \includegraphics[width=7cm]{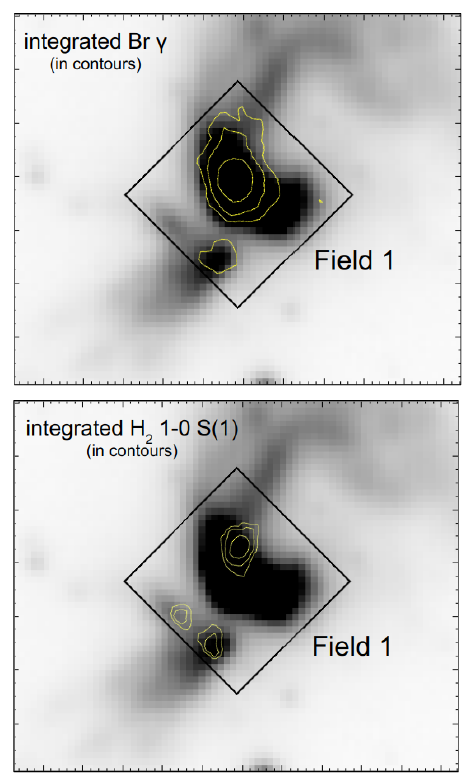}
    \caption{Integrated Br$\gamma$ and H$_{2}$ 1--0 S(1) emissions (continuum subtracted) towards Field\,1 (yellow contours) superimposed over the Ks image obtained with NIRI. The contours levels are 1.3, 2.1, and 4.0 $\times 10^{-19}$ erg s$^{-1}$cm$^{-2}$ for the Br$\gamma$ line, and 2.6, 3.0, and 3.8 $\times 10^{-19}$ erg s$^{-1}$cm$^{-2}$ for H$_{2}$ 1--0 S(1) line. In both cases the first contours is above 4$\sigma$. Box area is 3\s$\times$3\s.}
   \label{f1Integ}
\end{figure}

\begin{figure}[h]
    \centering
       \includegraphics[width=7cm]{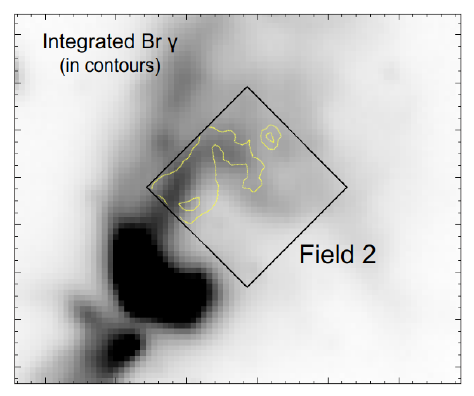}
    \caption{Integrated Br$\gamma$ emission (continuum subtracted) towards Field\,2 (yellow contours) superimposed over the Ks image obtained with NIRI. The contours levels are 0.7 and 1.2 $\times 10^{-19}$ erg s$^{-1}$cm$^{-2}$. First contour is above 4$\sigma$. Box area is 3\s$\times$3\s. }
   \label{f2Integ}
\end{figure}

\begin{figure}[h]
    \centering
       \includegraphics[width=8cm]{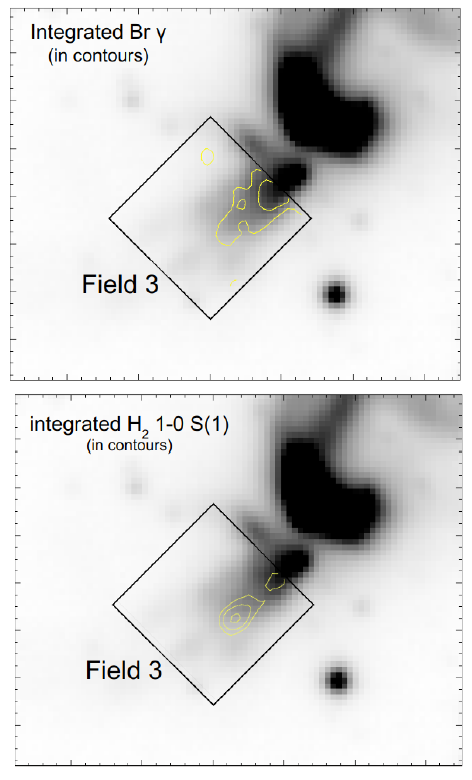}
    \caption{Integrated Br$\gamma$ and H$_{2}$ 1--0 S(1) emissions (continuum subtracted) towards Field\,3 (yellow contours) superimposed over the Ks image obtained with NIRI. The contours levels are 2.2 and 5.0 for the Br$\gamma$ line, and 0.6, 1.0, and 2.0 $\times 10^{-19}$ erg s$^{-1}$cm$^{-2}$ for the H$_{2}$ 1--0 S(1) line. In both cases the first contours is above 4$\sigma$. Box area is 3\s$\times$3\s. }
   \label{f3Integ}
\end{figure}

\begin{figure}[h]
    \centering
       \includegraphics[width=9cm]{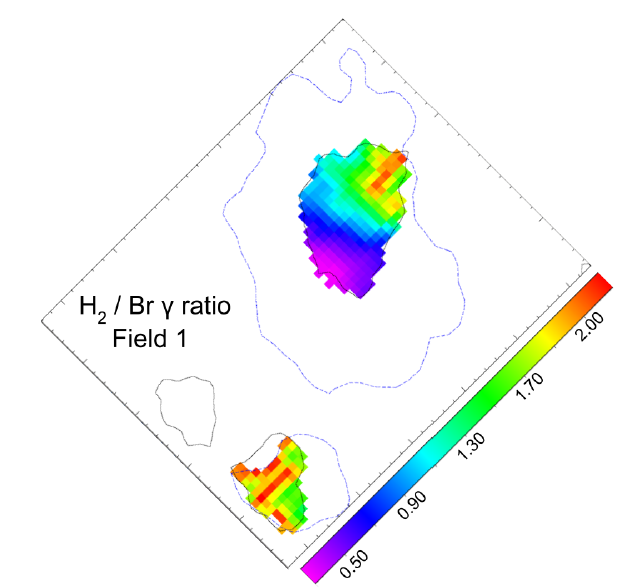}
    \caption{H$_{2}$/Br$\gamma$ ratio in Field\,1. Black (solid) and blue (dashed) contours are the lowest contours of the H$_{2}$ and Br$\gamma$ integrated emissions, respectively presented in 
    Fig.\,\ref{f1Integ}. These contours are displayed for reference.}
   \label{ratio}
\end{figure}

In this subsection, we report the emission lines detected from the NIFS observations towards each field (Fig.\,\ref{nifsObs}). Continuum maps are not shown here, as they do not present any significant difference from the broad-band Ks image obtained with NIRI in our previous work \citep{areal20}. 
After inspecting the fields along the whole observed spectral range (2.009 -- 2.435 $\mu$m), we only found Br$\gamma$ and H$_{2}$ S(1) 1--0 lines (rest wavelengths: 2.1661 and 2.1218 $\mu$m, respectively). Figure\,\ref{f1spect} displays, as an example, an averaged spectrum from Field\,1. Both lines were detected in Fields\,1 and 3, only Br$\gamma$ line appears in Field\,2, and no line is observed in Field\,4. Figures\,\ref{f1Integ}, \ref{f2Integ}, and \ref{f3Integ} exhibit maps of the observed integrated lines in each field. Figure\,\ref{ratio} presents the H$_{2}$/Br$\gamma$ ratio in Field\,1 at the regions in which both emissions are observed.

\begin{figure}[h!]
    \centering
       \includegraphics[width=8cm]{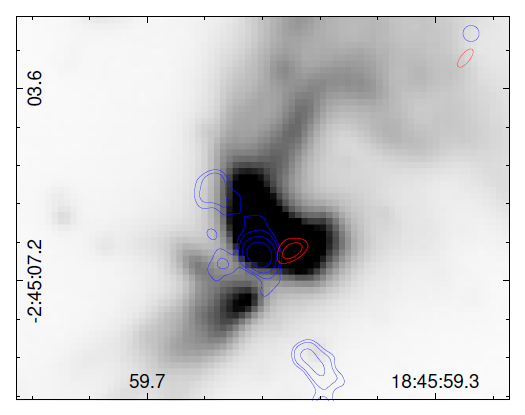}
    \caption{Ks image presented in our previous work obtained with NIRI at Gemini North \citep{areal20}. Blue contours represent the continuum emission at 1.3 mm obtained with ALMA at levels 1.8, 2.4, 4.0 and 8.0 mJy~beam$^{-1}$ (rms = 0.2 mJy~beam$^{-1}$). Radio continuum emission at 10 GHz acquired with JVLA is presented in red contours with levels of 20 and 50 $\mathrm{\mu Jy\,beam^{-1}}$ (rms = 6 $\mathrm{\mu Jy\,beam^{-1}}$). The beams of the ALMA and JVLA observations (blue and red) are presented at the top right corner.  }
   \label{vla+}
\end{figure}

\subsection{Discovering a compact radio continuum source}
\label{res_radio}

In our new radio continuum observations at $10~\mathrm{GHz}$ we detect a source located at R.A.$=$18:45:59.521, Dec.$=-$02:45:06.618. Figure\,\ref{vla+} 
displays in red contours the radio continuum emission at levels of 20 and 50~$\mathrm{\mu Jy\,beam^{-1}}$. The radio emission appears as a compact source located $\sim0\farcs7$ from the peak position of the central core detected at 1.3 mm (displayed in blue contours). The astrometry accuracy of JVLA and ALMA observations (see Sect.\,\ref{obs}) confirms that this is indeed a physical separation. The flux density integrated within the $50~\mathrm{\mu Jy\,beam^{-1}}$ contour level is $0.13\pm 0.01~\mathrm{mJy}$.  The complete radio image is shown in Appendix\,\ref{app0}.

\subsection{The dust emission and the molecular environment} 
\label{secc_molec}

As mentioned above, Fig.\,\ref{vla+} shows, in blue contours, the ALMA continuum emission at 1.3 mm depicting the dust cores in the region. It is worth noting that these ALMA data provide a better resolution of such cores compared to our previous works \citep{areal20,martinez24}. The main core is the only one with a rich abundance of complex molecules. In the others, some molecules are observed, but this is mostly due to extended emission that it is centred on the main core. This indicates that these cores are colder, very probably lacking of protostars within them. A chemical and physical characterization of the main dust core is included in Appendix\,\ref{app2}.

\begin{figure}[h!]
    \centering
           \includegraphics[width=8cm]{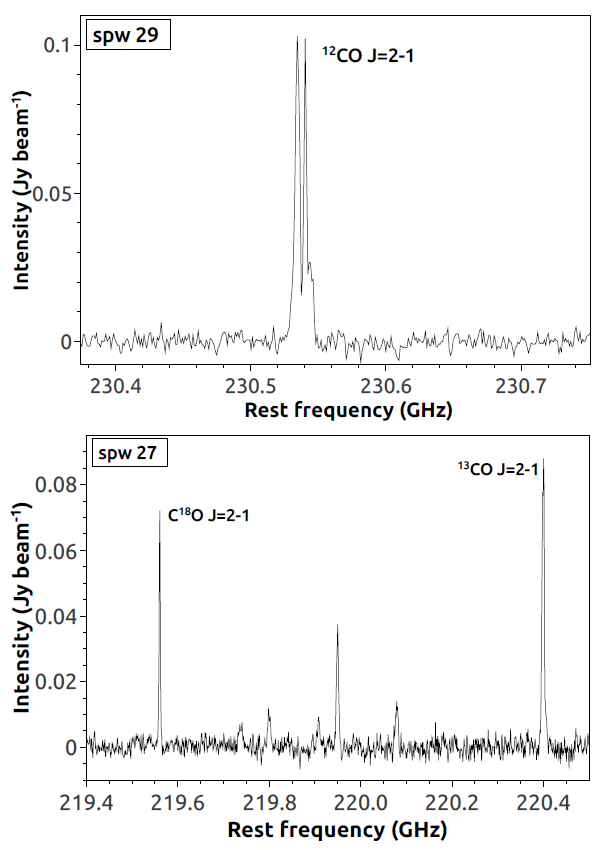}
    \caption{Portions of spectra from the spectral windows spw29 (up) and spw27 (bottom) obtained from a beam centred
    at the peak of the main dust core shown in Fig.\,\ref{vla+}. The lines of the $^{12}$CO (upper spectrum) and C$^{18}$O and $^{13}$CO (bottom spectrum) are indicated.}
   \label{spectra}
\end{figure}

We inspected the four spectral windows (spw) from the ALMA data cubes. Besides many lines of complex organic molecules (COMs) peaking mainly at the main core (see Appendix\,\ref{app2}), we found intense emission
of $^{12}$CO, $^{13}$CO, and C$^{18}$O  J=2--1 line (see Fig.\,\ref{spectra}). The systemic velocity of 101 \ks~for the molecular core was confirmed from the C$^{18}$O line. Since carbon monoxide and its isotopes effectively trace outflow activity and core envelopes, our main molecular analysis is focused on them. 

\begin{figure}[h!]
    \centering
           \includegraphics[width=8cm]{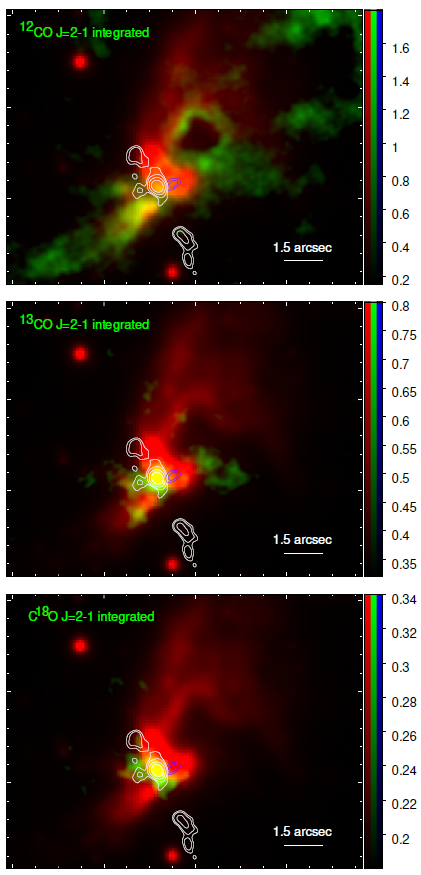}        
    \caption{Two-colours maps of YSO-G29 region. In green is displayed the \2, \3, and \8 J=2--1 emissions integrated along the whole frequency/velocity range in which each line extends as presented in Fig.\,\ref{spectra}. The Ks emission obtained with NIRI-Gemini is shown in red. White contours represent the continuum emission at 1.3 mm (same levels as Fig.\,\ref{vla+}), and blue contours are the continuum emission at 10 GHz (same levels as Fig.\,\ref{vla+}). The units of the colour bars at the right of each image are Jy beam$^{-1}$ \ks~and correspond to the green colour of the maps. The rms noise level of the molecular integrated emission is about 0.15 Jy beam$^{-1}$ \ks~for the three maps. }
   \label{mom0}
\end{figure}

To appreciate the morphology of the molecular gas in the region, as observed in carbon monoxide, we present in Fig.\,\ref{mom0} maps of the integrated emission (moment 0) of the three CO isotopes (in green) superimposed to the Ks NIRI-Gemini emission (in red). Additionally, contours of the continuum emissions at 1.3 mm and 10 GHz are included (white and blue, respectively). The \2 emission appears quite extended, likely depicting molecular outflows towards the south, and an intriguing shell-like feature towards the northwest. The emission of \3 concentrates primarily at the main 1.3 mm core with some extended structures close to the peak of the near-IR emission. The \8 emission is centralized at the main core with some emission nearby. 

Given the remarkable \2 morphology observed northwestwards of the main near-IR structure—a complete shell-like feature of 2\s~in diameter—, we present in Figs.\,\ref{cans1} and \ref{cans2} channel maps of the \2 emission every 1.2 \ks~(the spectral resolution of the observations). These channel maps show that the molecular shell-like structure extends from 87.6 to 96.0 \ks, forming a complete and closed shell perfectly surrounded by the near-IR emission. Indeed, we can wonder whether the near-IR morphology is due to such a molecular structure, or vice versa.
The velocity range, from 97.2 to 102.0 \ks, corresponding to the central and self-absorbed component of the \2 line towards the densest region, was excluded from the analysis. Finally, from 103.2 to 108.0 \ks~another conspicuous molecular structure appears towards the southeast. It seems to be a lobe, likely of a molecular outflow. In the first channels (103.2 and 104.4 \ks), it appears open and composed of two structures, while from channel 105.6 to 108.0 \ks~it appears collimated and quite aligned with the radio continuum source observed at 10 GHz (blue contours in the channel maps) in coincidence with the southeastern near-IR feature.

\begin{figure*}[tt]
    \centering
           \includegraphics[width=18cm]{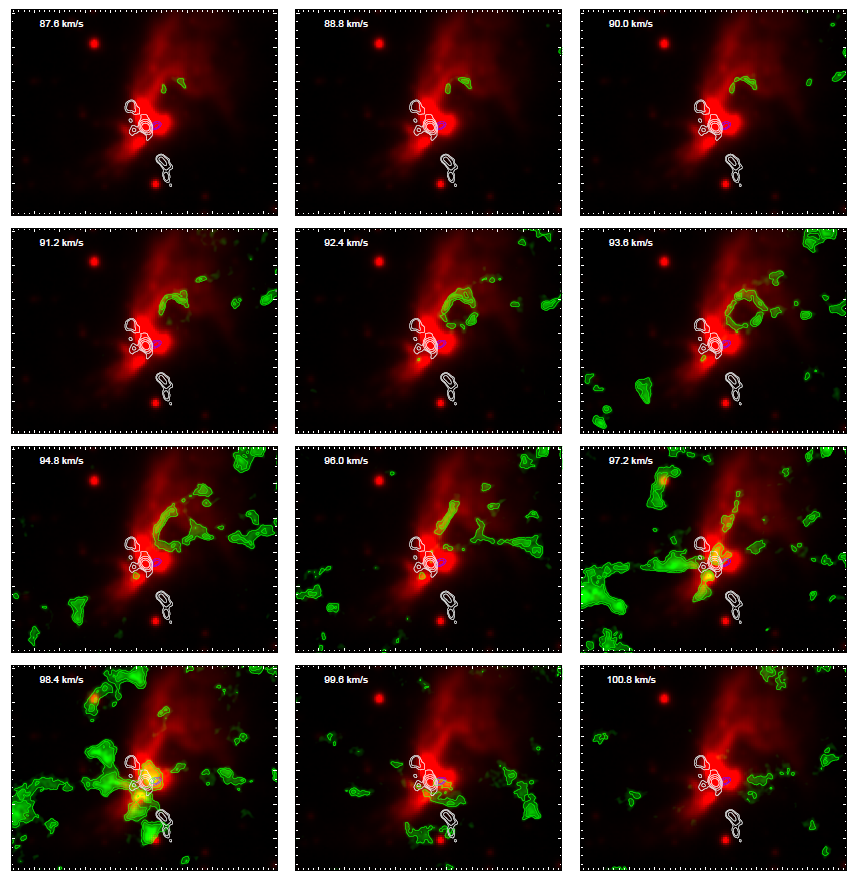}
    \caption{\2 J=2--1 channel maps towards YSO-G29 in green contours with levels of 0.075, 0.100, and 0.150 Jy beam$^{-1}$. Ks emission from NIRI-Gemini is displayed in red. Contours of the continuum emissions at 1.3 mm (white) and at 10 GHz (blue) are included (same levels as the above figures). Maps continue in Fig.\,\ref{cans2}. The rms noise level of each \2 channel is 0.025 Jy beam$^{-1}$. The systemic velocity of the source is 101 \ks. }
   \label{cans1}
\end{figure*}

\begin{figure*}[tt]
    \centering
           \includegraphics[width=18cm]{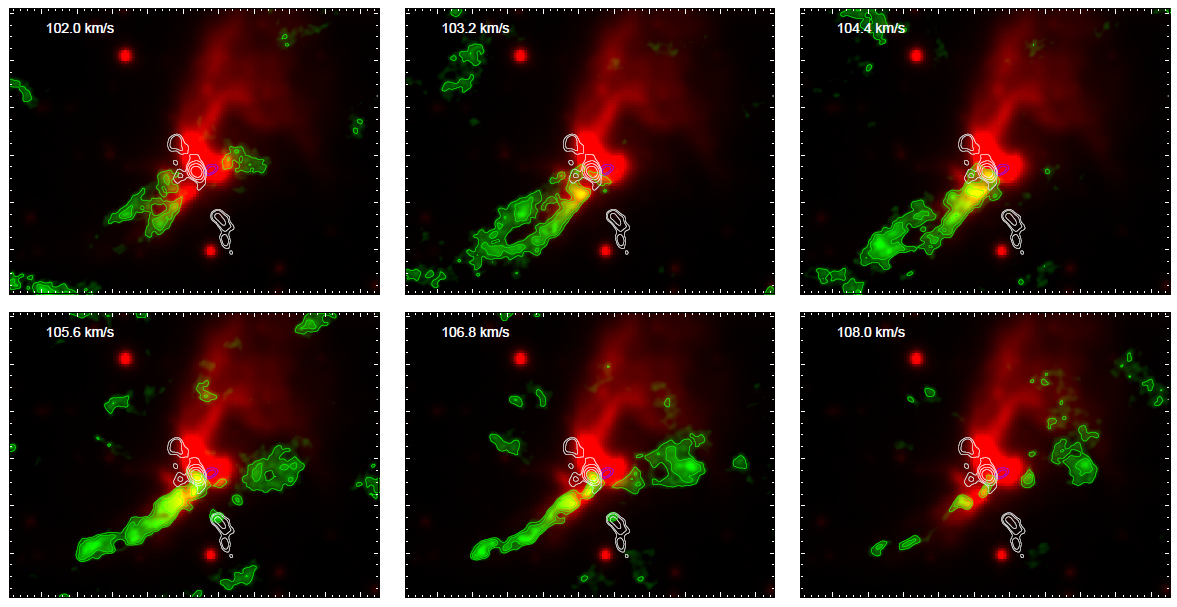}
 \caption{Continuation of Fig.\,\ref{cans1}. }
   \label{cans2}
\end{figure*}

\begin{figure}[h]
    \centering
         \includegraphics[width=8cm]{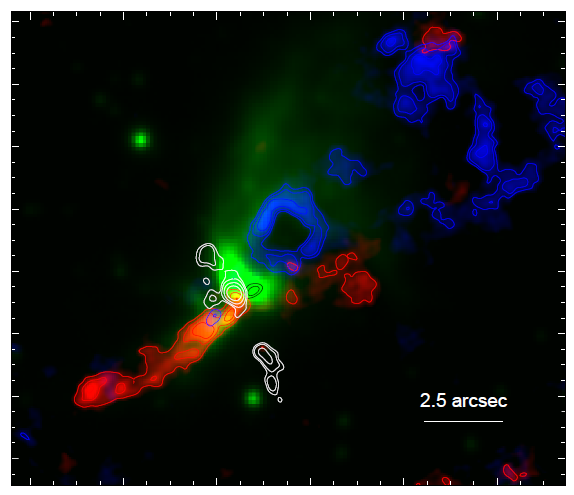}
\caption{\2 J=2--1 integrated in the velocity ranges 87.6--96.0 and 102--108 \ks~displayed in blue and red, respectively. The NIRI near-IR emission is displayed in green. The contour levels are 0.37, 0.50, and 0.75 Jy beam$^{-1}$ \ks~for the blue emission and 0.50, 0.75, and 1.00 Jy beam$^{-1}$ \ks~for the red emission. The white and black contours are the continuum emission at 1.3 mm and 10 GHz as presented in previous figures.  }
   \label{outflowsRB}
\end{figure}

To better appreciate red- and blue-shifted molecular gas associated with YSO-G29 we present in Fig.\,\ref{outflowsRB} the \2 emission integrated between 87.6 and 96.0 \ks~(blue) and between 102 and 108 \ks~(red) superimposed over the near-IR emission displayed in green.

\begin{figure}[h]
    \centering
         \includegraphics[width=8cm]{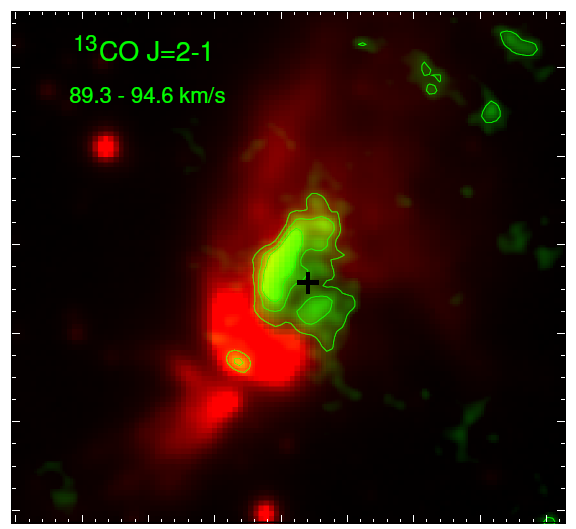} 
 \caption{\3 J=2--1 integrated between 89.3 and 94.6 \ks~with contours with levels of 0.095, 0.150, and 0.200 Jy beam$^{-1}$ \ks. The rms noise level of the integrated map is 0.030 Jy beam$^{-1}$ \ks. The black cross represents the position from which the \2 spectrum shown in Fig.\,\ref{12coabs} was extracted. }
   \label{13corange}
\end{figure}

\begin{figure}[h]
    \centering
         \includegraphics[width=8cm]{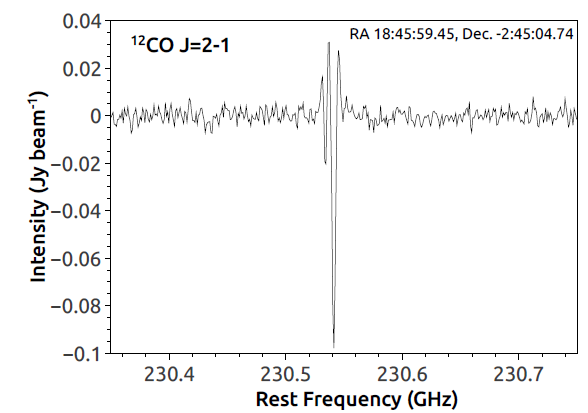} 
 \caption{\2 J=2--1 spectrum obtained towards the region indicated with the black cross in Fig.\,\ref{13corange} (coordinates are indicated at the top right corner). }
   \label{12coabs}
\end{figure}

Additionally, we further analyse the \3 emission in the area where the previously described \2 shell-like structure is observed. Figure\,\ref{13corange} displays the \3 integrated emission between 89.3 and 94.6 \ks, missed in the image that presents the whole integration (Fig.\,\ref{mom0}, medium panel). This demonstrates that the observed \2 shell is mostly filled with \3 emission. Finally, a \2 spectrum towards the centre of this region was extracted (black cross in Fig.\,\ref{13corange}). It is presented in Fig.\,\ref{12coabs}. These results cast doubt on the nature of the observed molecular shell in \2, implying it might result from self-absorption or flux loss in the line signal. This will be addressed in Sect.\,\ref{discus}. 

After confirming that the southeastern CO structure does not exhibit line absorption issues, we estimated the mass of the southern molecular outflow shown in Fig.\,\ref{mom0} (upper panel), in the channel maps in Fig.\,\ref{cans2} (channels 102 to 108 \ks), and in Fig.\,\ref{outflowsRB} (displayed in red). We obtained the CO column density from the CO J=2--1 transition following the works of \citet{turner1991} and \citet{miao1995}. From these studies the following equation was derived:

\begin{equation}
{\rm N_{tot}={\rm \left(\frac{2.04 \times 10^{20}}{\theta_a \theta_b}\frac{W}{g_kg_l{\nu_0}^3S_{ul}{\mu_0}^2}\right) \times \exp{\left(\frac{E_u}{kT_{ex}}\right)} \times Q_{rot} }},   
\label{RD} 
\end{equation}

\noindent where ${\rm \theta_a}$ and ${\rm \theta_b}$ (in arcsec) are the major and minor axes of the clean beam, respectively, W (in Jy beam$^{-1}$ \ks) is the integrated intensity, ${\rm g_{k}}$ is the K-ladder degeneracy, ${\rm g_{l}}$ is the degeneracy due to the nuclear spin, ${\rm \nu_{0}}$ (in GHz) is the rest frequency of the transition, ${\rm S_{ul}}$ is the line strength of the transition, $\mu_{0}$ (in Debye) is the permanent dipole moment of the molecule, E$_{\rm u}$/k is energy of the upper level, and T$_{\rm ex}$ is the excitation temperature.

From the moment zero map of the outflow structure, we estimated an integrated intensity of about 0.5 Jy beam$^{-1}$ \ks. Thus, assuming a T$_{\rm ex}$ = 20 K, and using the transition and molecular parameters obtained from the Splatalogue Database for Astronomical Spectroscopy (SDAS)\footnote{https://splatalogue.online/\#/advanced}, we derived an average CO column density of about $6.7 \times 10^{16}$ cm$^{-2}$. 

We estimated the total mass of the molecular outflow using the following equation:

\begin{equation}
    {\rm M_{out}= N(^{12}CO)~[H_{2}/CO]~\mu_{H_2}~m_{H}~A_{pixel}~N_{pixel}}
\end{equation}

\noindent where N($^{12}$CO) is the average column density of the lobe, ${\rm [CO/H_{2}]=10^{-4}}$ is the abundance ratio between the molecules, ${\rm \mu_{H_2}=2.72}$ is the mean molecular weight, ${\rm m_{H}=1.67 \times 10^{-24}}$~g is the mass of the hydrogen atom, ${\rm A_{pixel}}$ is the pixel area, and ${\rm N_{pixel}}$ is the number of pixels that fills the lobe. Using the distance of 6.2 kpc to YSO-G29, we obtained a mass for the outflow of about 0.2 \msol. 

Then the outflow rate ($\dot{M} = $ M$_{out}$/$t_{dyn}$), momentum ($P$ $=$ M $\bar{v}$) and  outflow momentum rate (${\dot{P}}= P/t_{dyn}$) were estimated. Dynamical time ($t_{dyn}$) is obtained from L/$v_{\rm max}$, where L is the length of the outflow (about 0.06 pc), and $\bar{v}$ and $v_{\rm max}$ are the median and maximum velocity with respect to the systemic velocity. The following values were obtained: $t_{dyn} \sim 8.3 \times 10^{3}$ yr, $\dot{M} = 2.4 \times 10^{5}$ \msol~yr$^{-1}$,  $P \sim 0.8$ \msol~\ks, and ${\dot{P}} \sim 1 \times 10^{-4}$ \msol~\ks~yr$^{-1}$. Caution is advised when interpreting these values due to potential projection effects. The small velocity range would indicate that such an outflow is observed mainly along the plane of the sky. In that case, the derived values for $P$ and ${\dot{P}}$ should be considered lower limits.

\section{Discussion}
\label{discus}

With a new set of high-quality data at higher angular resolutions than previous studies, we propose a compelling new scenario for the intriguing morphology of YSO-G29. This provides valuable observational insights into star formation processes. 

\subsection{Near-IR emission: lines from Gemini-NIFS}
\label{nirdisc}

The new Gemini-NIFS near-IR data reveal H$_{2}$ 1--0 S(1) and B$\gamma$ emission towards the main near-IR structure and the southern smaller feature (Figs.\,\ref{f1Integ} and\,\ref{f3Integ}, respectively) previously observed with Gemini-NIRI. On the other hand, only the Br$\gamma$ feature is detected to the north (Fig.\,\ref{f2Integ}). No other lines at near-IR were observed. 

The hydrogen Br$\gamma$ emission line is commonly observed
towards massive YSOs (e.g., \citealt{cooper13}).  Here, the extended
emission observed in Fields 1, 2, and 3, which has a morphological correspondence with the
continuum emission strongly suggests that this
line arises from stellar strong winds \citep{fedriani19} probably extending 
along cavities cleared out by jets. 

Conversely, it is well known that the excitation of H$_{2}$ 1--0 S(1) can be due to collisions or UV fluorescence. Ratios among different H$_2$ near-IR lines are typically used to distinguish between
excitation mechanisms (e.g.,\,\citealt{paron22}). However, the 1–0 S(1) is the only H$_{2}$ line detected in our observations. By comparing the regions in which H$_{2}$ is detected with their corresponding Br$\gamma$ emission in Field\,1, we find that the H$_{2}$ 1--0 S(1)/Br$\gamma$ ratio presents a gradient going from values $<1$ to somewhat larger than the unity at the main region of near-IR emission (see Fig.\,\ref{ratio}). This suggests a combination of excitation mechanisms for the H$_{2}$ emission in the region: UV excitation (ratios $<1$) and shocked gas (ratios $>1$) (see \citealt{hatch05,chen15,reiter24}). For its part, the southern near-IR structure in Field\,1 presents H$_{2}$/Br$\gamma$ ratios larger than 1, suggesting that this H$_{2}$ feature is due to shocked gas.

We conclude that the detected near-IR lines are compatible with excitation and ionization from UV radiation propagating in a highly perturbed ambient by the activity of YSO-G29, with some contribution of shocked gas. 
It should be emphasized that no line at near-IR was detected towards the dark lane (see Fig.\,\ref{present}) purportedly to be a disc or a toroid of material observed edge-on as suggested in \citet{areal20}. As spectral lines can be detected towards discs and/or toroids (e.g., \citealt{murakawa13}), we discard such a nature for the dark lane, proposing it to be just a region lacking near-IR emission.

\subsection{Molecular outflows}
\label{secc_outflow}

Using ALMA data in band 6, with an angular resolution of 0\farcs3, we discovered molecular outflows associated with YSO-G29 (systemic velocity about 101 \ks) at small spatial scales (resolving structures of about 1800 au). This discovery also challenges the conclusions drawn in \citet{areal20} regarding the red- and blue-shifted jets/outflows. 

We discovered a conspicuous \2 outflow extending southeastwards (see Figs.\,\ref{mom0}, \ref{cans2}, and \ref{outflowsRB}). Considering the velocity range over which this outflow extends, although not particularly broad, we infer that it is a redshifted outflow. Hence, the features observed northwestwards can be attributed to blueshifted gas (see Figs.\,\ref{cans1} and \ref{outflowsRB}), tracing a more open and fragmented structure. Changing the scenario regarding the directions of jets and outflows proposed in \citet{areal20}, we conclude that, at present, the observed extensions of the near-IR features make sense: the northern-southern asymmetry in the extension of the structures observed at the Ks broad-band partially should be due to possible extinction effects that are more pronounced in the southern red-shifted feature.

The southern CO outflow appears quite collimated, somewhat clumpy along the line of sight. For instance, in velocity channels at 102.0, 103.2, and 104.4 \ks~(see Fig.\,\ref{cans2}), it can be appreciated that the outflow seems to be composed of two structures: one more aligned with the main millimetre core and the other along the same direction of the radio continuum source. Then, in velocity channels at 105.6 and 106.8 \ks, the outflow presents a well-collimated structure aligned mostly towards the position of the radio continuum source. Notably, in these velocity channels, the CO emission seems to originate at the position of the main millimetre core, extending southwards before bending to the southeast, where it forms the collimated outflow previously mentioned. 

The obtained mass of such a southern molecular outflow, about 0.2 \msol, is in agreement with the mass of many CO outflows associated with cores embedded in massive 70 $\mu$m dark clumps \citep{li2020}. It is worth mentioning that the authors carried out a statistical study of molecular outflows using ALMA observations based on the \2 J=2--1 line, and they found a median outflow velocity of about 21 \ks~with respect to the systemic velocity. Therefore, the low velocity of the YSO-G29 southern outflow, about 4 \ks, indicates that this molecular feature should extend mostly along the plane of the sky, confirming that $P$ and ${\dot{P}}$ should be lower limits. Moreover, one must also be careful with the parameters obtained, as the analysed outflow may be the product of outflows from more than one source. 

Regarding the northern/northwestern molecular ambient, there does not appear to be any collimated feature as in the southern region. On the contrary, we found fragmented molecular gas with the most remarkable features being a quite extended CO shell-like structure with an excellent morphological correspondence with the near-IR emission from Gemini-NIRI (see Fig.\,\ref{mom0} upper panel, channels from 87.6 to 94.8 \ks~in Fig.\,\ref{cans1}, and Fig.\,\ref{outflowsRB}). However, the presence of a \3 structure contained by such a \2 shell-like structure (Fig.\,\ref{13corange}), also perfectly matching with the near-IR emission, casts doubt on the shell nature of the molecular gas. Indeed, the \2 spectra towards the centre of this structure are highly absorbed (Fig.\,\ref{12coabs}). This can be due to high optical depth effects that usually \2 lines suffer in dense gaseous structures. Additionally, the deep \2 absorption spectral feature could indicate missing flux coming from more extended spatial scales that are filtered out by the interferometer. Certainly, this phenomenon can produce not only missing flux but also deep absorption features (e.g., \citealt{choi04,rodon12,paron21}). We conclude that the observed structure is not a molecular shell enclosed by near-IR emission but a non-collimated molecular outflow component, likely produced by one or multiple ejections from YSO-G29 that have impacted the feature observed at near-IR.

The near-IR emission related to YSOs at these small spatial scales usually has a morphology of cone-shaped nebulae. They are cavities cleared out in the circumstellar material by the action of jets and winds, which are bright mainly at near-IR \citep{reip00,bik05,bik06,paron13,paron16,sanna19}. Considering that it could be the case of YSO-G29, taking into account the perfect morphological correspondence between the molecular emission and the intriguing morphology of the northern near-IR feature, we suggest that it could be due to an ejection of molecular gas that has disrupted the cone-shaped cavity previously carved (this will be further discussed in Sect.\,\ref{escenario}).

\subsection{Radio continuum source}
\label{radiosource}

Weak and compact radio sources in star-forming regions may be associated with \hii~regions in their earlier stages of evolution (namely, HC/UC \hii~regions), where thermal radio emission forms as a consequence of photoionization by a recent born star. Alternatively, radio emission may be associated with jets of YSOs, where thermal free–free emission arises as a consequence of shock ionization in the wind/jet of the protostars. In some cases, these so-called ionised jets trace the base of the jet (i.e., they trace the position of the powering protostar), while in other cases they form in knots along the jet, away from the protostar. The most powerful knots, where strong shocks arise, can also be sources of non-thermal radio emission of synchrotron origin, sometimes refereed to as non-thermal lobes (see \citealt{anglada18} for a review of radio jets in YSOs).

In an attempt to classified our discovered compact radio source, we compare the radio luminosity ($S_\nu D^2$) and the bolometric luminosity ($L_{bol}$). \cite{anglada92} found that $S_\nu D^2$ of ionised jets is positively correlated with $L_{bol}$, and subsequent surveys in the radio band have confirmed that this correlation is valid for both low- and high-mass protostars (see \citealt{purser16,rosero19,purser21}). 
For this purpose, we refer to the work of \cite{purser21}. We extrapolate our radio flux at $10\,\mathrm{GHz}$ to $5.8\,\mathrm{GHz}$ assuming some spectral behaviour and estimate the radio luminosity for a distance $D=6.2\,\mathrm{kpc}$. For an ionised jet/knot, we take a canonical spectral index $\alpha=0.6$, while for a HC/UC \hii~region, we use $\alpha=-0.1$, which is the theoretical value of a photoionised nebula in the optically thin regime. We obtain $S_{5.8}D^2=3.5$ and $5.2\,\mathrm{mJy\,kpc^2}$ for $\alpha=0.6$ and $-0.1$, respectively.
Regarding the bolometric luminosity of the region, we estimated $L_{bol}\sim 3.7\times 10^4~\mathrm{L_\odot}$ following the method presented in \cite{rosero19} (see Appendix\,\ref{appSed}). 

In Fig.\,\ref{purser}, we show the $S_{5.8}D^2$ vs. $L_{bol}$ diagram of \cite{purser21} (see their Figure 4), where we have included our compact radio source (yellow dot crossed by dashed lines)\footnote{For easiness of comparison, we use a radio luminosity of $4\,\mathrm{mJy\,kpc^2}$ and $L_{bol}=4\times10^4~\mathrm{L_\odot}$.}.
The figure shows that the compact radio source is located in the region of the diagram populated by ionised jets and non-thermal lobes from high-mass protostars. However, taking into account the large dispersion of the diagram and the possibility that multiple sources in the region are contributing to $L_{bol}$, we cannot discard that the compact radio source may be a HC/UC \hii~region. Indeed, if $\sim 3.7\times 10^4~\mathrm{L_\odot}$ is an upper limit to the bolometric luminosity of the radio source, its actual position in the diagram may be displaced in the direction of the Lyman continuum of a ZAMS star.   
Considering the possibility of an \hii~region, we can estimate the Lyman continuum photon flux of the putative star responsible of the ionisation. From \citet{kurtz94}, using the radio flux at 10 GHz ($S_\nu = 86~\mu\mathrm{Jy}$), $T_e = 10^{4}$ K and the parameter $a=1$, we derived $N_{Ly} = 3.2\times 10^{44}$ ph s$^{-1}$. Following \citet{avedisova79}, this amount of Lyman photons should be produced by a B1 type star.

\begin{figure}[h]
    \centering
         \includegraphics[width=9cm]{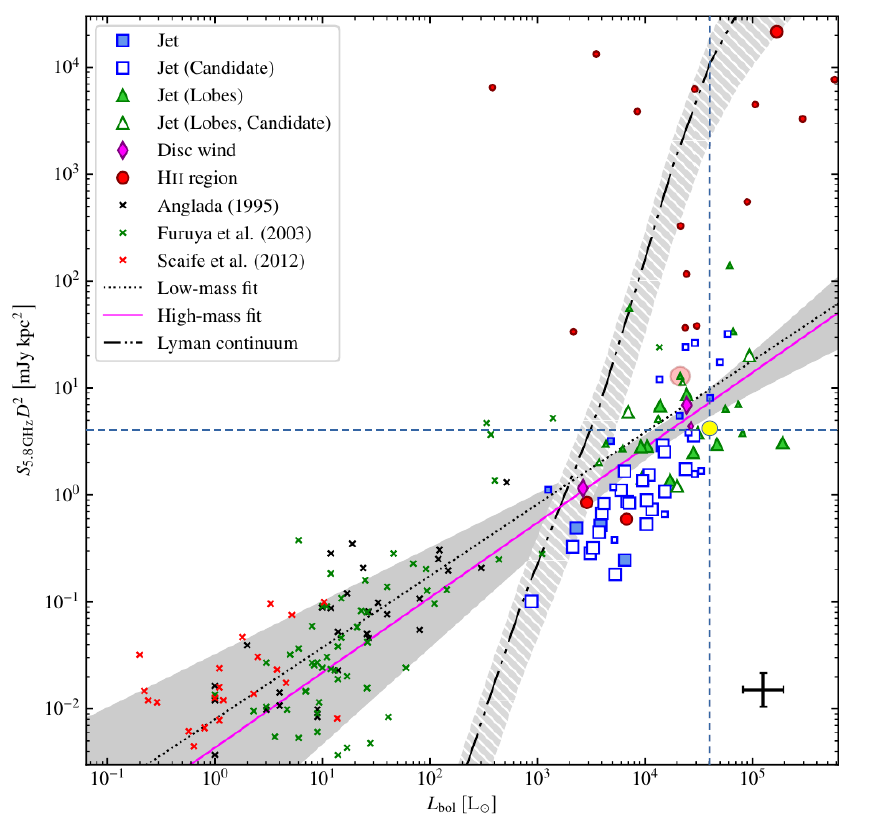} 
\caption{Reproduction of Fig.\,4 from \cite{purser21}: radio luminosity ($S_\nu d^2$) vs. bolometric luminosity ($L_{bol}$) in which our source is included (yellow circle crossed by dashed lines).
} 
   \label{purser}
\end{figure}

In summary, our analysis of the available information about the radio compact source points to single-source and multiple-source scenarios for the YSO-G29 region. In the former, the compact radio source may be a thermal knot or a non-thermal lobe located along the jet of a putative protostar likely situated at the centre of the 1.3 mm core (the unique core with high temperatures in the region). In the multiple-source scenario, the ALMA main core may be hosting a massive protostar while the radio source may be indicating the position of a second massive young star: either a massive protostar powering an ionised thermal jet (which traces the base of the jet, i.e. the position of the powering protostar), or a more evolved YSO, an HC/UC \hii~region excited by a B1 type star. Conducting a radio spectral index study would provide valuable insights for discerning the nature of the radio source. Based on these possibilities, the following section explores the most plausible interpretation for this region, focusing on the complex morphology observed mainly at near-IR and the molecular emission distribution.

\subsection{Proposed scenario for YSO-G29}
\label{escenario}

As shown above, the discovered radio continuum emission could be an UC\hii~region, an ionised jet from a massive object, or even a non-thermal lobe. This last possibility would suggest that it should be a jet coming from an undetected object embedded in the main molecular core, a similar case as reported for instance in \citet{beltran16}. However, taking into account the complexity of the morphology of the different emissions, hereafter we will favour a scenario of multiple YSOs in the region. 

Given that the radio continuum source is not associated with a peak of a molecular core (traced by dust emission), we suggest that it is more likely to be an UC\hii~region than an ionised jet from a massive object. Such an UC\hii~region would probably have evacuated the surrounding dust. Then, taking into account that the main 1.3 mm core is the only one in the region with a rich presence of COMs peaking at its maximum (it has indeed high temperatures; see Appendix\,\ref{app2}), we propose that this core and the discovered radio continuum source are the sole objects related to active star formation in the region. Hence, the positional offset between them merits further discussion. 

Assuming the main core must contain an internal heat source, we propose the existence of a protostar, or a non-detected massive star, embedded within it. Considering the last case, from the rms noise level of the radio continuum image, we can roughly estimate an upper limit of Lyman photons emanating from such a putative star, and then estimate an upper limit for its spectral type. Following the same procedure as described above for the compact radio source, we obtain an amount of ionising photons of N$_{\rm Ly} = 2.2 \times 10^{43}$ ph s$^{-1}$ from considering a flux density equals to the rms noise level of the radio image. Then, based on \citet{avedisova79}, we conclude that the spectral type of the putative star could be B3 or later, i.e., a star with lower, or much lower mass than the star responsible of the detected radio compact source. 
Of course, the object embedded in this core could be a young protostar that has not begun to ionise the surrounding gas yet. 

In addition, it is worth noting that the Br$\gamma$ emission detected at near-IR with Gemini-NIFS (see Sect.\,\ref{nirdisc}) shows the 
presence of ionised gas in the region, reinforcing the hypothesis of ionising sources, which indeed could be young \hii~regions. Even though the Br$\gamma$ peak is not
coincident with the position of the radio continuum source proposed to be an UC\hii~region, it may indicate ionised gas escaping from it, with some contribution from the source embedded in the dust main core. In conclusion, a crucial finding of this study is the identification of a likely YSO binary system.

It is known that binary and multiple systems are a common outcome of the star formation process (\citealt{ricciardi25}, and references therein). YSO-G29 could be a binary system composed of a star later than B3, or a protostar, non-ionizing yet, and a B1 type star generating an UC\hii~region. If this is the case, the observed system would present a projected separation of 0.02 pc (about 4000 au) between both stellar components. Following \citet{ricciardi25}, this could be a probable separation for a wide binary system, likely generated by turbulent fragmentation of a molecular cloud. If this is the case, assuming both stars began forming at the same time, and considering that massive stars form more rapidly, the mass difference between the two objects is consistent with the less evolved stage of the embedded one. Additionally, it is important to note that if the embedded source in the core is a protostar, it is likely that it will not become a massive star (see the estimated mass for the core in Appendix\,\ref{app2}). 

The scenario of a wide binary system would explain the morphology of the different features detected in infrared and molecular lines. For instance, the observed structures of the molecular outflows (see Figs.\,\ref{cans1} and \ref{cans2}, and Sect.\,\ref{secc_outflow}) would be due to ejections originating from both objects. What is indeed outstanding is the perfect morphological matching between the molecular gas and the near-IR emission towards the northwest (see Figs.\ref{cans1}, \ref{outflowsRB} and \ref{13corange}). Given that it is usual to detect cone-like near-IR features towards YSOs tracing cavities cleared out in the circumstellar material by the action of jets and winds \citep{reip00,bik05,bik06,paron13,paron16,sanna19}, we are probably observing such a type of cavity, but disrupted at the western part. 

Thus, it is probable that one of the sources generated the mentioned cavity, and that molecular outflows from the other source subsequently disrupted it. This proposed scenario could finally elucidate the infrared structure associated with YSO-G29, which has remained virtually unexplained since our initial study focusing on this region was published in 2020 \citep{areal20}.

This type of interaction might be relatively common in star-forming regions with multiple components (see \citealt{bally16}, and references therein).  For instance, more recently, \citet{zapata18} reveal the collision between two outflows in a protostar binary system. Our case resembles this scenario, in which it is occurring the disruption of a YSO circumstellar cavity by an outflow. These kinds of interactions and disruptions are important phenomena for accounting when studying massive star-forming regions with high angular resolutions.

\section{Conclusion}

Using a high-quality new set of multiwavelength data, we pointed to shed light on YSO G29.862-0.0044 (YSO-G29). The most intriguing aspect was the morphology that the YSO presents in near-IR, which was observed more than five years ago with Gemini-NIRI. We conclude that the results presented in this work may resolve this question, and they can provide valuable observational evidence for our understanding of star-forming regions that exhibit high confusion.
 
From the discovery of a compact radio source and the analysis of a hot molecular core with a probable stellar object embedded within it, we found evidence of a YSO binary system, in which one of its components may have generated a cavity in the surrounding interstellar medium. 
Our most significant result is the observational evidence that such a cavity was partially disrupted by a molecular outflow likely generated by the other component of the system. This would be a similar case to others reported in the literature regarding collisions of outflows and cavities disruptions.

These results, obtained from highly detailed observations of a distant star-forming site (about 6.2 kpc), not only help to complete the picture of what is happening in this particular region, but also reveal a phenomenon that should be considered when investigating high-mass star-forming regions that exhibit high confusion.

\begin{acknowledgements}

We thank the anonymous referee for her/his insightful comments and suggestions that helped us to improve this work.
N.C.M. is a doctoral fellow of CONICET, Argentina.
This work was partially supported by the Argentine grants PIP 2021 11220200100012 and PICT 2021-GRF-TII-00061 awarded by CONICET and ANPCYT.
This work is based on the following ALMA data: ADS/JAO.ALMA $\#$ 2021.1.00311.S. ALMA is a partnership of ESO (representing its member states), NSF (USA) and NINS (Japan), together with NRC (Canada), MOST and ASIAA (Taiwan), and KASI (Republic of Korea), in cooperation with the Republic of Chile. The Joint ALMA Observatory is operated by ESO, AUI/NRAO and NAOJ.

\end{acknowledgements}

%
%

\bibliographystyle{aa}  
\bibliography{ref}
\IfFileExists{\jobname.bbl}{}
{\typeout{}
\typeout{****************************************************}
\typeout{****************************************************}
\typeout{** Please run "bibtex \jobname" to optain}
\typeout{** the bibliography and then re-run LaTeX}
\typeout{** twice to fix the references!}
\typeout{****************************************************}
\typeout{****************************************************}
\typeout{}
}

\begin{appendix}

\section{Radio image}
\label{app0}

\begin{figure}[h]
    \centering
         \includegraphics[width=9cm]{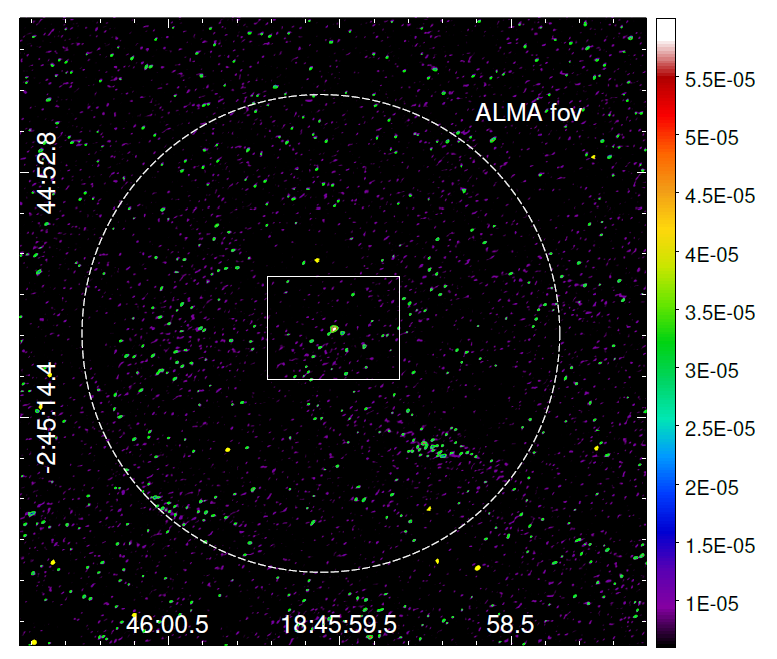} 
 \caption{Radio continuum emission at $10~\mathrm{GHz}$. The white circle is ALMA's fov and the white rectangle is the zoomed-in region shown in Fig. \ref{radio_img2}. Color scale is expressed in $\mathrm{Jy\,beam^{-1}}$ and goes from 6 to $60~\mathrm{\mu Jy\,beam^{-1}}$, corresponding to 1 and 10 times the image noise ($\sigma_{rms}=6~\mathrm{\mu Jy\,beam^{-1}}$), respectively. Countor levels are: $-3\sigma_{rms}$ (yellow), $2\sigma_{rms}$ (green), and $3\sigma_{rms}$ (red). 
}
 \label{radio_img}
\end{figure}

\begin{figure}[h]
    \centering
         \includegraphics[width=9cm]{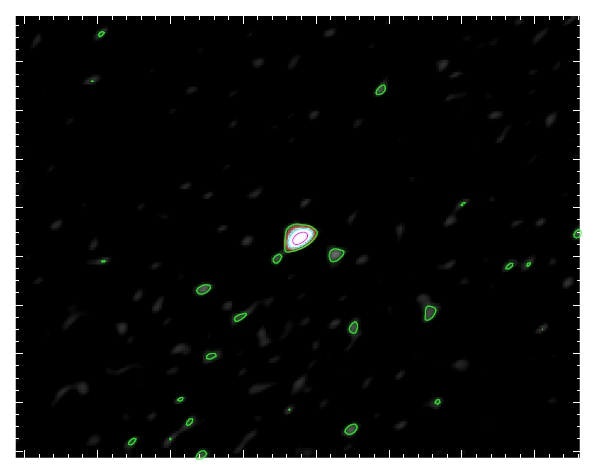} 
 \caption{Zoomed-in view of the compact radio source. Contour levels are: $2\sigma_{rms}$ (green), $3\sigma_{rms}$ (red), $5\sigma_{rms}$ (cyan), and $10\sigma_{rms}$ (magenta).
}
 \label{radio_img2}
\end{figure}

In Fig.\,\ref{radio_img}, we show the JVLA radio continuum image at $10~\mathrm{GHz}$, covering the field-of-view (fov) of ALMA observation. 
Figure \ref{radio_img2} shows a zoomed-in image of the compact radio source.

\section{Bolometric luminosity of the region}
\label{appSed}

\begin{figure}[h]
    \centering
         \includegraphics[width=9.5cm]{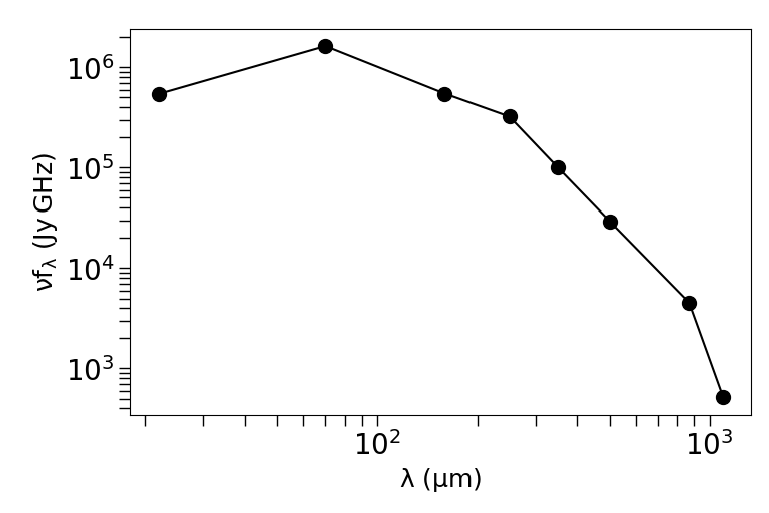} 
 \caption{Spectral energy distribution in the mid-IR, submillimeter and millimeter bands in the direction of YSO-G29. The bolometric flux of the region is estimated as the area under the linear interpolation of the fluxes. }
   \label{SED_bolo}
\end{figure}

We estimated the bolometric luminosity of the region following the same procedure of R19, who used a linear interpolation of the fluxes between the mid-IR and the millimetre bands. In Table \ref{tabla_bolo} we report the fluxes $f_\lambda$ of the catalogued sources from the Bolocam Galactic Plane Survey (BGPS, \citealt{rosol10}), the APEX Telescope Large Area Survey of the Galaxy (ATLASGAL, \citealt{urqu14}), and the Herschel Space Observatory infrared Galactic Plane Survey (Hi-GAL, \citealt{moli16}). In the mid-IR band, R19 used the $24~\mathrm{\mu m}$ flux from the MIPSGAL survey, but in our case the catalogued source MIPSGAL MG029.8569-00.0434 overlaps with a masked out area of the MIPSGAL image, probably due to saturation. Instead, we used the flux at $22~\mathrm{\mu m}$ of the Wide-field Infrared Survey Explorer (allWISE, \citealt{cutri14}). In Fig.\,\ref{SED_bolo} we show the spectral energy distribution and the linear interpolation. 
The area under the interpolation gives the bolometric flux $A = (3.06 \pm 0.01)\times 10^{-8}~\mathrm{erg\,s^{-1}\,cm^{-2}}$. To estimate the error, we obtained a lower and an upper limits of the area by interpolating $f_\lambda-\Delta f_\lambda$ and $f_\lambda+\Delta f_\lambda$, respectively, where $\Delta f_\lambda$ are the fluxes errors of Table \ref{tabla_bolo}. 
For a distance $d=6.2\pm0.6~\mathrm{kpc}$, the bolometric luminosity is $L_{bol} = 4\pi Ad^2 = (3.7\pm0.7) \times 10^{4}~\mathrm{L_{\odot}}$.

\begin{table}[h]
\centering
\caption{Mid-IR, submillimetre and millimetre fluxes $f_\lambda$ in the direction of YSO-G29.}
\label{tabla_bolo}
\begin{tabular}{lcr}
\hline
\hline
Source              &   Band &   $f_{\lambda}\pm \Delta f_\lambda$ \\         
                         & [$\mathrm{\mu m}$] &      [Jy]      \\
BGPS G029.863$-$00.048   &   1100    &        $1.90\pm0.2$   \\
AGAL 029.862$-$00.044    &    870    &       $13.1\pm 2.2$    \\
HIGALPL029.8647$-$0.0450 &    500    &       $49.1\pm 2.0$   \\
HIGALPM029.8643$-$0.0447 &    350    &       $117.4\pm2.1$   \\ 
HIGALPS029.8636$-$0.0451 &    250    &       $270.0\pm1.4$   \\
HIGALPR029.8632$-$0.0437 &    160     &      $292.7\pm1.2$   \\
HIGALPB029.8624$-$0.0437 &     70   &        $382.2\pm0.7$   \\
allWISE J184559.51$-$024506.1 &  22 &        $40.0\pm0.1$     \\
                  
\hline
\end{tabular}
\tablefoot{The WISE $22~\mathrm{\mu m}$ flux ($f_{22}$) was calculated from the magnitude $m_{22}=-1.70\pm0.03$ of the allWISE Catalog, using the formula $f_{22}=f_010^{-2.5/m_{22}}$, with the zero point $f_0=8.36~\mathrm{Jy}$ reported in the VOSA SED Analyzer \citep{bayo08}. }

\end{table}

\section{Chemical and physical characterization of the main dust core}
\label{app2}

To complement the multi-wavelength characterization of YSO-G29, we examined the chemical composition of the main dust core revealed by the ALMA observations, the only one in the region with noticeable emission of COMs. Hot molecular cores like G29 generally have elevated temperatures, high densities, and a rich molecular content. The extreme conditions cause the ice on the dust grains to sublimate, releasing molecules into the gas phase and initiating complex chemical reactions. Molecular composition studies of these environments provide insight into both the physical conditions of the gas and the formation pathways of complex organic molecules.

\begin{figure}
    \centering
    \includegraphics[width=9cm]{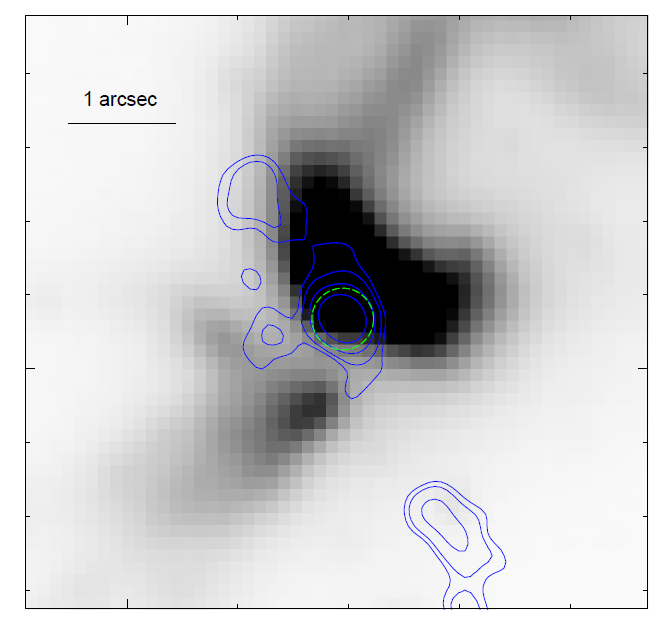}
    \caption{Ks Gemini-NIRI image with ALMA continuum at 1.3 mm contours as presented above. The green dashed circle indicates the position from which spectra presented in Figs.\,\ref{spectraCOMs} and \ref{spectraCOMs2} were extracted.}
    \label{region}
\end{figure}

In this context, we conducted a detailed spectral analysis of the ALMA Band 6 data to identify the molecular species in the hot core associated with YSO-G29. This study integrates spectral data from two independent datasets: Project 2021.1.00311.S (PI: Liu, Sheng-Yuan), mentioned in Sect.\,\ref{alma data}, and Project 2015.1.01312.S (PI: G. Fuller), described in \cite{ortega23}. These observations provide complementary coverage of the 216.6–242.7 GHz frequency range, distributed across eight spectral windows.
Spectra were extracted from them using a circular region centred at the continuum emission peak (see green dashed circle in Fig.\,\ref{region}). We visually inspected each spectrum to identify molecular lines and labelled all distinguishable features. Molecular identification was performed using the SDAS, considering the rest frequency, line strength, upper-level energy, and molecular species. This process led to identifying a chemically rich inventory displayed in Figs.\,\ref{spectraCOMs} and \ref{spectraCOMs2}. It includes both simple and complex species such as CH$_{3}$CN, HC$_{3}$N, H$_{2}$CO, HNCO, CN, SO, SO$_{2}$, and various isotopologs, among others. 

The rotational diagram analysis was carried out for certain detected molecules $-$CH$_{3}$CN, CH$_{3}$CCH, E-CH$_{3}$OH, A-CH$_{3}$OH, HC$_{3}$N, and CH$_{3}$OCH$_{3}$$-$ that exhibit at least two unblended transitions accompanied by well-characterized upper-level energies (E$_{\rm u}$) and integrated intensities. 
Following the formalism described in \cite{ortega23} (see their Sect.~4.3.1.) we computed the natural logarithm of the column density per statistical weight (ln(N$_{\rm u}$/g$_{\rm u}$)) as a function of the E$_{\rm u}$ (Fig.\,\ref{Trots}). This calculation used the integrated line intensities, beam dimensions, and spectroscopic parameters retrieved from the SDAS, which compiles spectroscopic constants from the Cologne Database for Molecular Spectroscopy (CDMS)  and Jet Propulsion Laboratory Molecular Spectroscopy (JPL) catalogs. Under the assumptions of local thermodynamic equilibrium (LTE) and optically thin emission, the inverse of the slope of the linear fit yields the rotational temperature (T$_{\rm rot}$). In this approximation T$_{\rm rot}$ is considered a proxy for the kinetic temperature (T$_{\rm k}$) of the gas.

The derived rotational temperatures span a wide range, from approximately 70 K to over 360 K, indicating that chemically distinct species trace gas layers with varying thermal conditions within the hot core.
The derived T$_{\rm rot}$ for CH$_{\rm 3}$CN and HC$_{\rm 3}$N are 361 K and 236 K, respectively. These values are consistent with those reported in previous studies of hot molecular cores. For instance, \cite{ortega23} found that CH$_{\rm 3}$CN traces gas layers with temperatures around 340 K, indicating its association with the innermost, hottest regions of the core. Similarly, \cite{chen25} reported rotational temperatures for HC$_{\rm 3}$N in the range of 160–335 K, with a mean value of 235 K, in excellent agreement with our result. However, it is important to note that their analysis was based on vibrationally excited transitions of the molecule, while in our case, only the ground state was considered. Although we did not detect vibrationally excited lines of this molecule, their presence in the region cannot be ruled out. Therefore, obtaining a similar temperature from ground-state transitions is plausible and consistent with the presence of compact warm gas. 
It should be noted that our HC$_{\rm 3}$N fit is based on only two transitions. Although the derived T$_{\rm rot}$ is consistent with previous high-resolution studies, additional data would be necessary to confirm this estimate with higher confidence.

The detection of vibrationally excited transitions of CH$_{\rm 3}$CN (v$_{8}=1$) and CH$_{3}$OH (v$_{\rm t}=1$) provides further evidence of hot, compact gas near the central source. These lines are typically excited in regions with temperatures above $\sim$ 200–300 K and have been observed in well-studied hot cores such as Orion \citep{sutton86} and in star-forming regions like G9.62$+$0.19 \citep{peng22}. These detections in YSO-G29 indicate the presence of deeply embedded, high-excitation gas layers, consistent with physical conditions near a massive protostar.

Both T$_{\rm rot}$ obtained for CH$_{3}$OH in their A and E configurations indicate that this molecule probes gas found at moderate temperatures in the core. A more detailed analysis of the relative behaviour of the A/E symmetry states and their implications for the thermal and chemical structure of the source is discussed in Martinez et al. (2025, in prep.). On the other hand, CH$_{3}$CCH and CH$_{3}$OCH$_{3}$ appear to trace colder gas components within the hot core environment. Both molecules exhibit lower rotational temperatures in our analysis, suggesting that their emission originates from more extended and less heated regions of the core. CH$_{3}$CCH, in particular, has been widely used as a tracer of lukewarm gas in dense molecular clouds and hot cores, often reflecting temperatures in the 30–100 K range \citep{santos22,ortega23}. Similarly, CH$_{3}$OCH$_{3}$ is typically associated with chemically evolved gas released into the gas phase through thermal desorption. However, it can persist in cooler layers, especially if it forms on grain surfaces and is desorbed early during the heating phase, where it is found to trace similar temperatures \citep{fontani07,li24}. The relatively low excitation temperatures derived for CH$_{3}$OCH$_{3}$ in YSO-G29 are therefore consistent with an origin in the outer and cooler regions of the hot core—likely within the envelope—where complex organic molecules may survive for longer periods before being further processed or destroyed. In addition, sulfur-bearing species such as SO, SO$_{2}$, and H$_{2}$CS were also identified, highlighting the chemical diversity of the region and suggesting active sulfur chemistry operating in this range of physical conditions.

In summary, the molecular inventory and excitation analysis of YSO-G29 reveal a chemically rich and thermally stratified hot core. The range of rotational temperatures derived from species probing different excitation regimes—ranging from compact, high-temperature gas to cooler, more extended components—points to the presence of a radial thermal gradient, as expected in internally heated protostellar environments. The detection of vibrationally excited transitions and complex organics further supports the advanced chemical evolution of the core. 

As a final step in the physical characterization of the chemically rich core, we estimated its mass using the approach described by \citet{kau08} (eq.\,B.1):

\begin{eqnarray}
{\rm M_{gas}=0.12~M_\odot \left[exp\left(\frac{1.439} {(\lambda/mm)(T_{dust}/10~K)}\right)-1\right]} \\ \nonumber {\rm \times\left(\frac{\kappa_{\nu}}{0.01~cm^2~g^{-1}}\right)^{-1}\left(\frac{S_{\nu}}{Jy}\right)\left(\frac{d}{100~pc}\right)^2\left(\frac{\lambda}{mm}\right)^3}
\label{mass}
\end{eqnarray}

\noindent We used the integrated flux at 1.3 mm, S$_{\nu} = 0.03$ Jy, and assumed a dust temperature, T$_{\rm dust} = 361$ K, derived from the CH$_3$CN emission and considering the thermal coupling of gas and dust. The dust opacity per gram of matter at 1.3 mm, $\kappa_{\nu}$, was adopted as 0.013~cm$^2$g$^{-1}$ \citep{lin2021}. The resulting mass is approximately 1 M$_\odot$, for a distance of 6.2 kpc.

\begin{figure*}
    \centering
    \includegraphics[width=17.3cm]{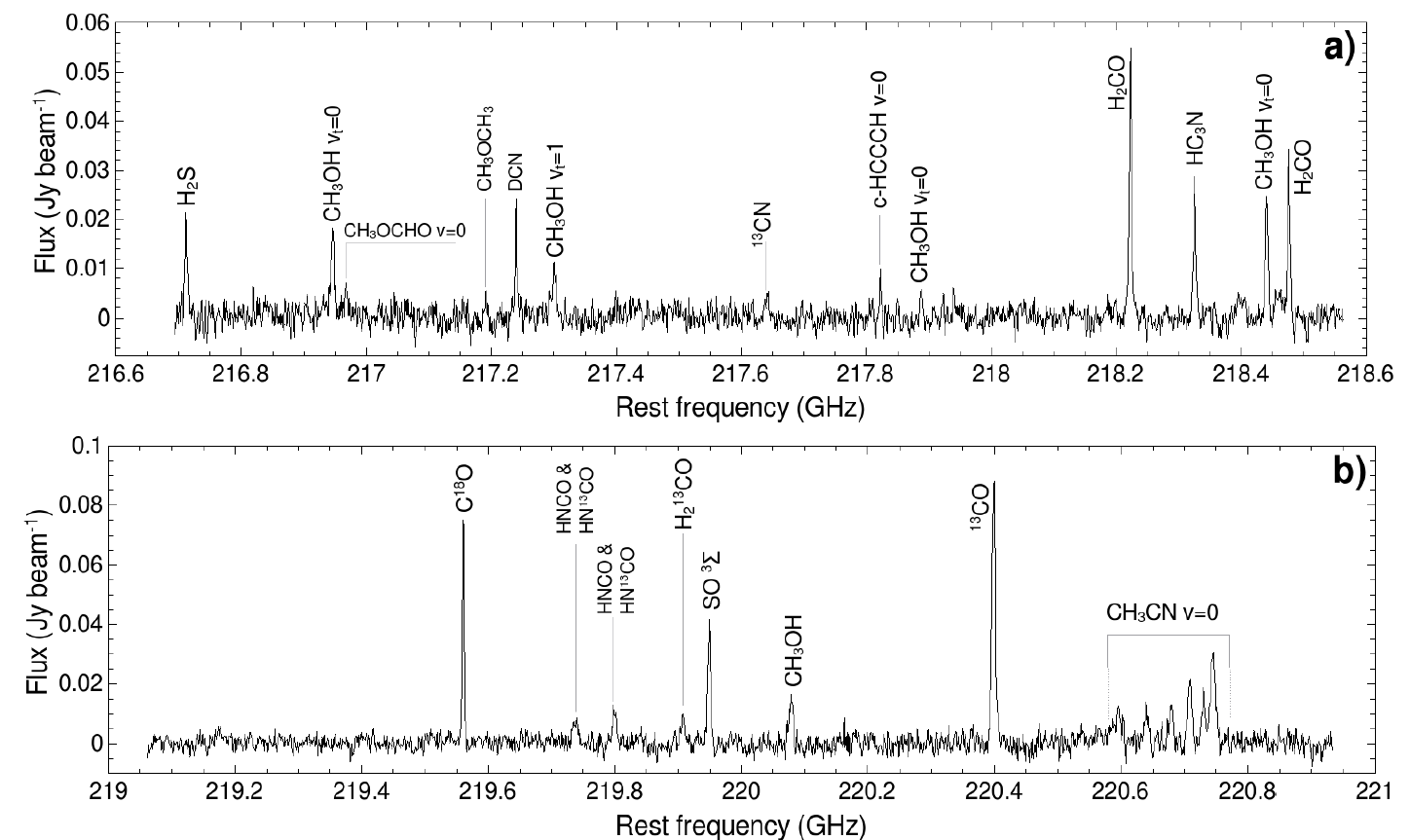}
    \includegraphics[width=17.4cm]{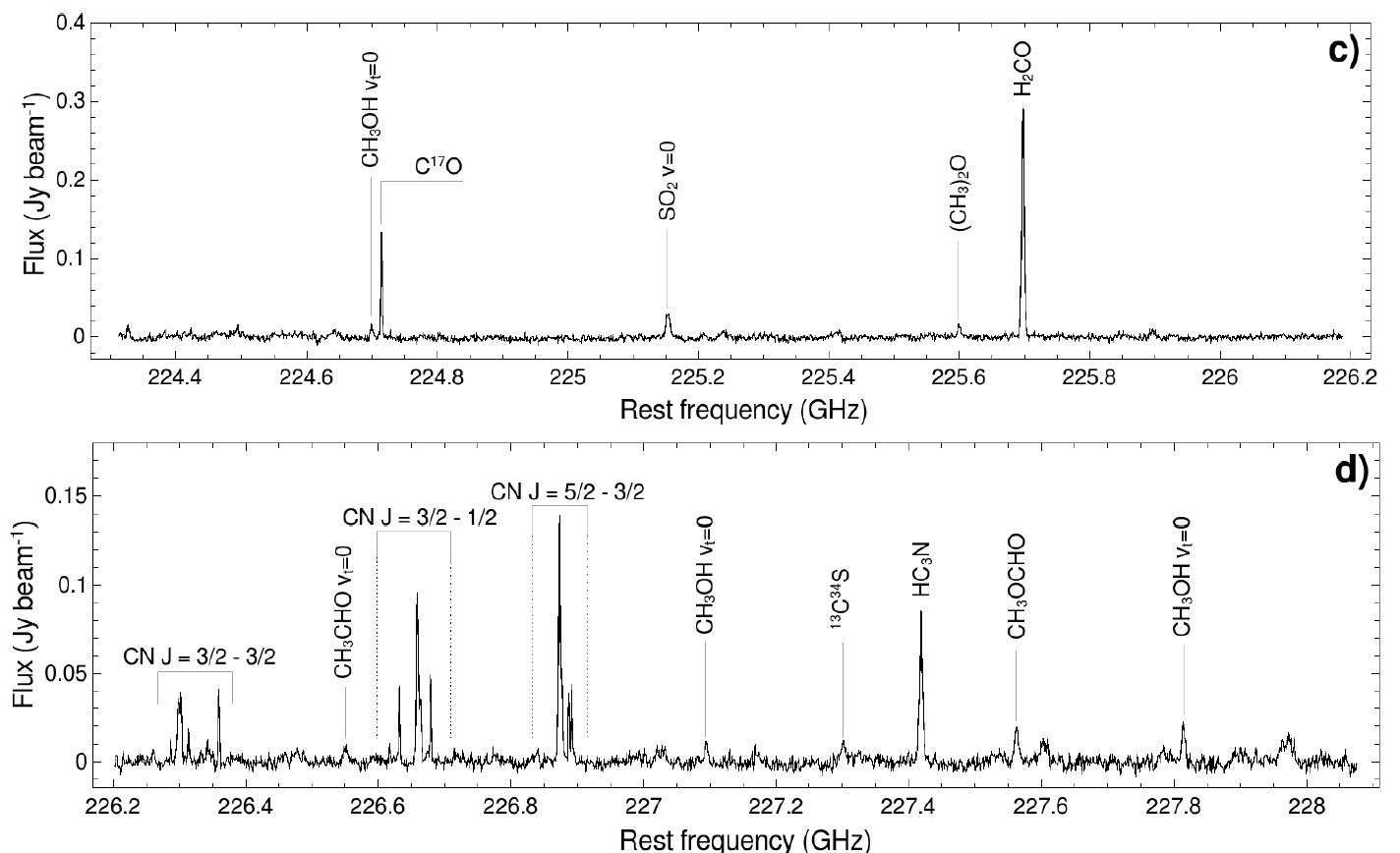}
    
    \caption{Spectra of the eight spectral windows analysed in this study. Panels a), b), e), and f) present data obtained from the observations reported by Liu et al. (Project 2021.1.00311.S), whereas panels c), d), g), and h) correspond to the dataset acquired by Fuller et al. (Project 2015.1.01312.S). Each spectrum was extracted from a beam-size region centered at the peak of the continuum emission. Molecular transitions were identified using the SDAS database.}
    \label{spectraCOMs}
\end{figure*}

\begin{figure*}
    \includegraphics[width=17.6cm]{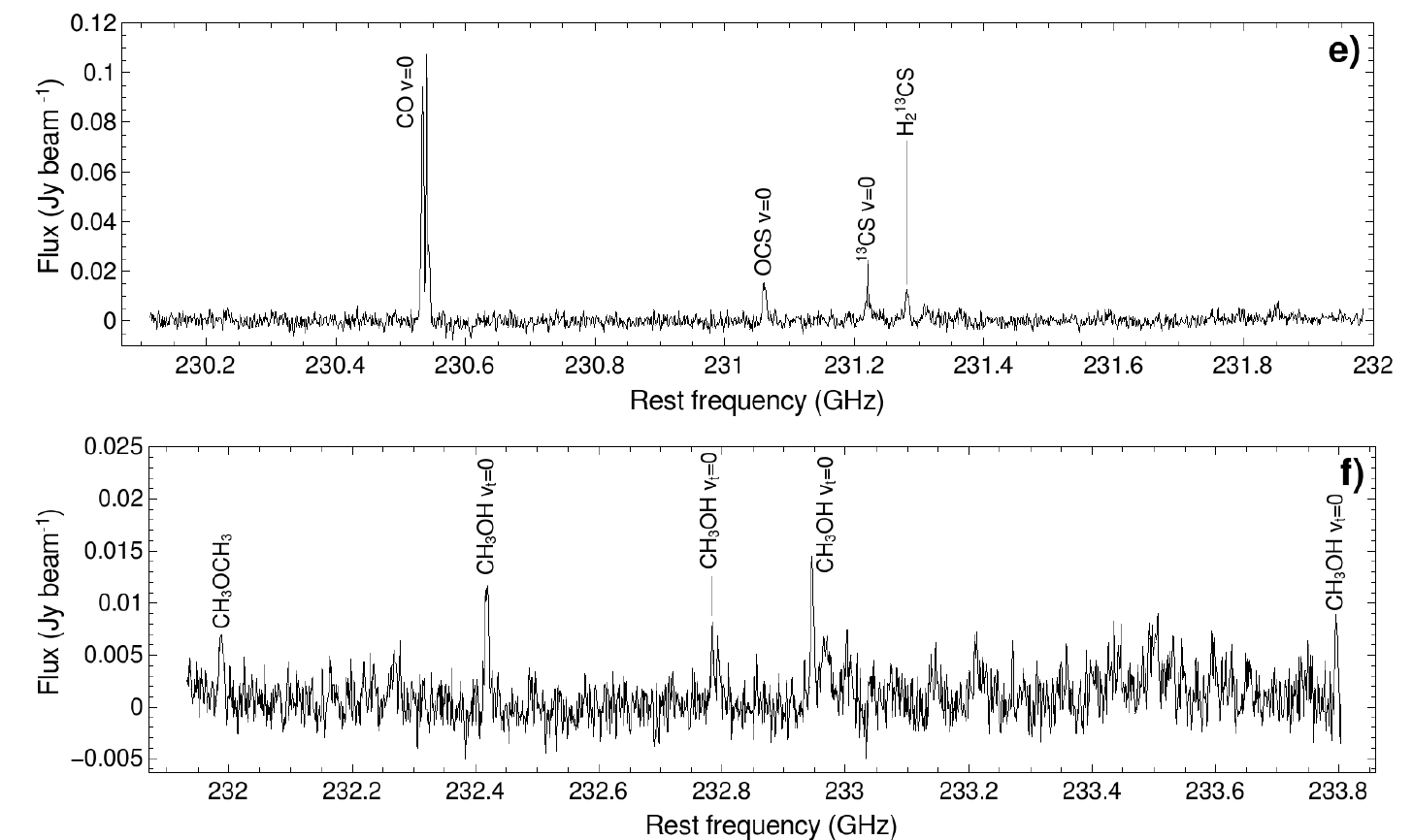}
    \includegraphics[width=17.4cm]{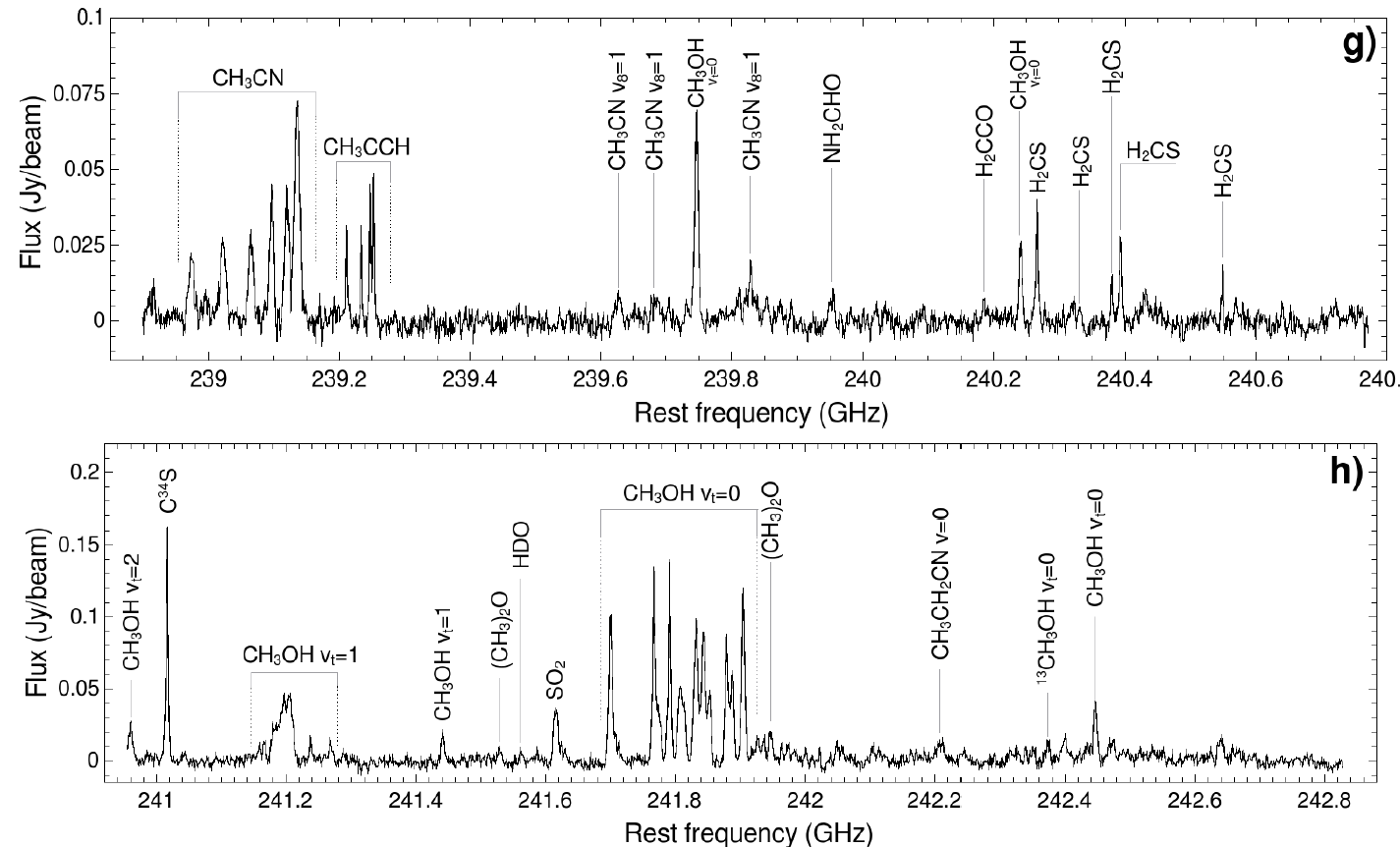}
    \caption{Continuation of Fig.\ref{spectraCOMs}.}
    \label{spectraCOMs2}
\end{figure*}

\begin{figure*} [h!]
    \centering
    \includegraphics[width=18cm]{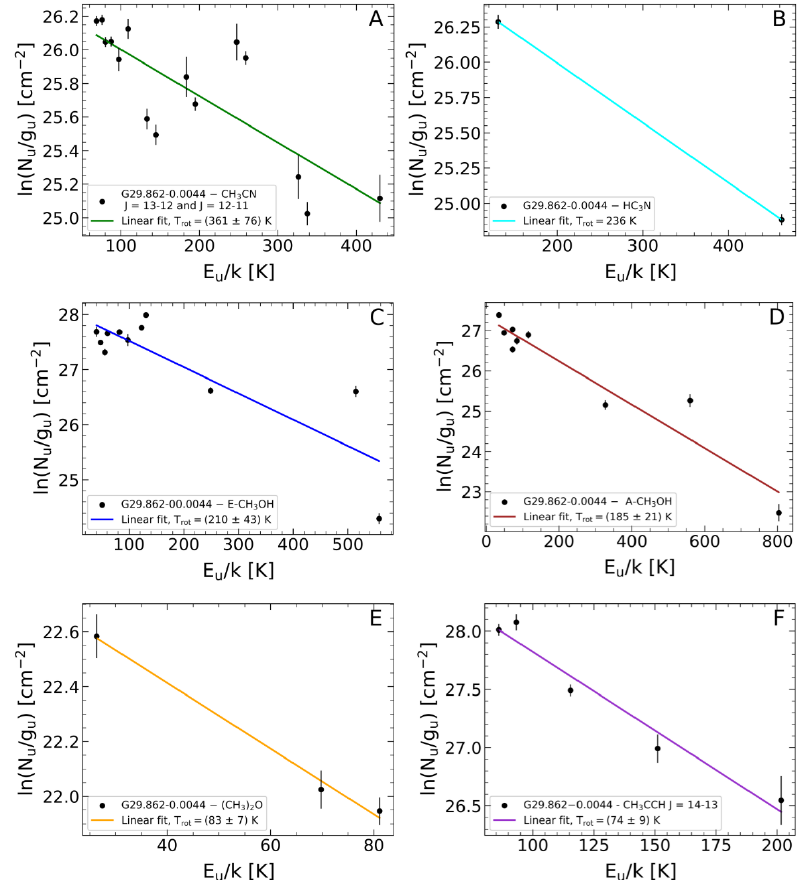}
    \caption{Rotational diagrams of selected molecules towards YSO-G29: (a) $\rm CH_{3}CN$, (b) $\rm HC_{3}N$, (c) $\rm E-CH_{3}OH$, (d) $\rm A-CH_{3}OH$, (e) $\rm CH_{3}OCH_{3}$, and (f) $\rm CH_{3}CCH$. Panels c), d), and f) are reproduced from Martinez et al. (2025, in prep.). The coloured lines represent the best linear fit of each dataset. For $\rm HC_{3}N$, the uncertainty of the rotational temperature was not estimated due to insufficient degrees of freedom.}
    \label{Trots}
\end{figure*}

\end{appendix}

\label{lastpage}

\end{document}